\documentclass[twocolumn,trackchanges]{aastex631}

\usepackage{amsmath} 
\usepackage{color}
\usepackage{soul}

\begin{document}

\title{The Obscured Fraction of Quasars at Cosmic Noon}

\correspondingauthor{Bovornpratch Vijarnwannaluk}
\email{bovornpratch.v@astr.tohoku.ac.jp}

\author[0000-0003-2213-7983]{Bovornpratch Vijarnwannaluk}
\affiliation{Astronomical Institute, Tohoku University, Aramaki, Aoba-ku, Sendai, Miyagi 980-8578, Japan}

\author[0000-0002-2651-1701]{Masayuki Akiyama}
\affiliation{Astronomical Institute, Tohoku University, Aramaki, Aoba-ku, Sendai, Miyagi 980-8578, Japan}

\author[0000-0001-7825-0075]{Malte Schramm}
\affiliation{Graduate school of Science and Engineering, Saitama University, 255 Shimo-Okubo, Sakura-ku, Saitama City, Saitama 338-8570, Japan}

\author[0000-0001-7821-6715]{Yoshihiro Ueda}
\affiliation{Department of Astronomy, Kyoto University, Kitashirakawa-Oiwake-cho, Sakyo-ku, Kyoto 606-8502, Japan}

\author[0000-0001-5063-0340]{Yoshiki Matsuoka}
\affiliation{Research Center for Space and Cosmic Evolution, Ehime University, 2-5 Bunkyo-cho, Matsuyama, Ehime 790-8577, Japan}

\author[0000-0002-3531-7863]{Yoshiki Toba}
\affiliation{National Astronomical Observatory of Japan, 2-21-1 Osawa, Mitaka, Tokyo 181-8588, Japan}

\author[0000-0002-7712-7857]{Marcin Sawicki}
\affiliation{Institute for Computational Astrophysics and Department of Astronomy and Physics, Saint Mary’s University, Halifax, NS B3H 3C3, Canada}

\author[0000-0001-8221-8406]{Stephen Gwyn}
\affiliation{Canadian Astronomy Data Centre, NRC-Herzberg, 5071 West Saanich Road, Victoria, British Columbia, V9E 2E7, Canada}

\author{Janek Pflugradt}
\affiliation{Astronomical Institute, Tohoku University, Aramaki, Aoba-ku, Sendai, Miyagi 980-8578, Japan}

\begin{abstract}

Statistical studies of X-ray selected Active Galactic Nuclei (AGN) indicate that the fraction of obscured AGN increases with increasing redshift, and the results suggest that a significant part of the accretion growth occurs behind obscuring material in the early universe. We investigate the obscured fraction of highly accreting X-ray AGN at around the peak epoch of supermassive black hole growth utilizing the wide and deep X-ray and optical/IR imaging datasets. A unique sample of luminous X-ray selected AGNs above $z>2$ was constructed by matching the XMM-SERVS X-ray point-source catalog with a PSF-convolved photometric catalog covering from $u^*$ to 4.5$\mu \mathrm{m}$ bands. Photometric redshift, hydrogen column density, and 2-10 keV AGN luminosity of the X-ray selected AGN candidates were estimated. Using the sample of 306 2-10 keV detected AGN at above redshift 2, we estimate the fraction of AGN with $\log N_{\rm H}\ (\rm cm^{-2})>22$, assuming parametric X-ray luminosity and absorption functions. The results suggest that $76_{-3}^{+4}\%$ of luminous quasars ($\log L_X\ (\rm erg\ s^{-1}) >44.5$) above redshift 2 are obscured. The fraction indicates an increased contribution of obscured accretion at high redshift than that in the local universe. We discuss the implications of the increasing obscured fraction with increasing redshift based on the AGN obscuration scenarios, which describe obscuration properties in the local universe. Both the obscured and unobscured $z>2$ AGN show a broad range of SEDs and morphology, which may reflect the broad variety of host galaxy properties and physical processes associated with the obscuration.

\end{abstract}


\section{Introduction} \label{sec:intro}

Observational results indicate that supermassive black holes (SMBH) exist ubiquitously in all massive galaxies (see review by \citealt{1995ARA&A..33..581K,2013ARA&A..51..511K}). However, how such massive black holes grow over cosmic time is not well understood. Active galactic nuclei (AGN) represent a key phase of SMBH growth during which SMBHs are actively accreting mass. The number density of AGN peaks at redshift 1 to 3, also known as the cosmic noon, and shows a trend in which the number density of more luminous AGN peaks earlier at redshift $\sim 3$ than that of less luminous AGN and then declines towards the local universe \citep{2014ApJ...786..104U,2014MNRAS.439.2736D,2015MNRAS.451.1892A}. This era represents a crucial period where the bulk of the cosmic SMBH mass density (90\%) was gained through mass accretion in luminous quasars \citep{1982MNRAS.200..115S,2004MNRAS.351..169M,2014MNRAS.439.2736D,2014ApJ...786..104U}. 

One large uncertainty in tracing the SMBH accretion growth is the fraction of obscured accretion. Studies using hard X-rays above 10 keV \citep{2009MNRAS.399..944M,2011ApJ...728...58B} and cosmic X-ray background synthesis studies have shown that a non-negligible fraction of AGN at low redshift are obscured \citep{1995A&A...296....1C,2003ApJ...598..886U,2006ApJ...639..740B, 2007A&A...463...79G,2009ApJ...696..110T}. 

Although optical imaging surveys have been successful in constructing large samples of quasars at high redshifts, identification of obscured AGN using optical color selection is challenging due to their similarity in color to galaxies with no ongoing AGN activity. Several emission line diagnostic diagrams were constructed to identify obscured AGN from galaxies \citep{1981PASP...93....5B,2016MNRAS.456.3354F}. However, the classification requires spectra, which can be time expensive to obtain in large numbers for faint objects at high redshifts. Alternatively, multiwavelength AGN signatures such as X-ray, mid-infrared, or radio emission can also be used to identify obscured accretion activity. Among the tracers of AGN activity at high redshift, hard X-ray datasets (E$>$2keV) currently provide the most reliable indicator of AGN activity. In addition, they provide the most complete view of the high redshift AGN population compared to other AGN selection methods thanks to the strong contrast against stellar light and the lower bias against obscuration.

In studies at low redshifts ($\rm z<2$), the obscured fraction shows a clear anti-correlation with AGN luminosity where more luminous quasars are less likely to be obscured. This trend has been observed ubiquitously in various AGN samples selected by optical emission lines, X-rays, and mid-infrared emission \citep{2005MNRAS.360..565S,2005ApJ...635..864L,2007A&A...468..979M,2008A&A...490..905H,2011ApJ...728...58B,2013PASJ...65..113T,2021ApJ...912...91T}. Following the orientation-based AGN unification scheme \citep{1993ARA&A..31..473A, 1995PASP..107..803U}, the trend can be explained by the inner torus structure receding outwards due to strong illumination and sublimation of dust, thus the opening angle within which the central engine is directly observable increases with luminosity or in other words, the dust covering factor decreases \citep{1991MNRAS.252..586L,2014ApJ...788...45T}. Another possible scenario is that the obscured fraction is controlled by radiation pressure on dust particles: AGN blow out the obscuring material after exceeding an $N_{\rm H}$-dependent critical Eddington ratio \citep{2006MNRAS.373L..16F,2008MNRAS.385L..43F,2009MNRAS.394L..89F}.  Evidence supporting this scenario is found in AGN in the local universe showing that the nuclear column density depends on the Eddington ratio and AGNs whose accretion activity exceeds the critical effective Eddington ratio tend to show strong blowout winds \citep{2009MNRAS.394L..89F,2017Natur.549..488R, 2019MNRAS.489.3073B, 2021ApJS..257...61Y,2021arXiv210614527T}. 

High resolution hydrodynamic simulations demonstrate that a torus-like structure is a natural outcome of gas accretion towards the nuclear region and that the torus properties are closely related to the SMBH and nuclear ISM properties \citep{2012MNRAS.420..320H,2012ApJ...758...66W,2012ApJ...759...36R,2016ApJ...828L..19W,2016MNRAS.458..816H}. These simulations can reproduce the observed column density distribution of AGN in the local universe as well as the obscured fraction by assuming that the torus is clumpy and that AGN feedback via radiation pressure clears some portions of sight lines \citep{2012MNRAS.420..320H,2012ApJ...758...66W,2012ApJ...759...36R,2016MNRAS.458..816H,2016ApJ...828L..19W}.

There is another trend that among Compton-thin AGN (CTN-AGN) with $\log N_{\rm H}\ (\rm cm^{-2}) < 24$ the fraction of obscured AGN with $\log N_{\rm H}\ (\rm cm^{-2})>22$ increases from the local universe up to redshift 2 \citep{2003ApJ...598..886U,2005ApJ...635..864L,2006ApJ...639..740B,2006ApJ...652L..79T,2008A&A...490..905H,2009ApJ...696..110T,2014ApJ...786..104U,2015MNRAS.451.1892A,2015ApJ...802...89B}. Since most of the obscuring material is thought to be concentrated in the nuclear region \citep{2018ARA&A..56..625H}, the redshift dependence suggests that the nuclear region contains a larger amount of gas analogous to the larger gas fraction in galaxies at high redshifts than those in the local universe \citep{2010Natur.463..781T,2013ARA&A..51..105C}. Alternatively, the evolution of the obscured fraction among CTN-AGN may be driven by the X-ray obscuration from the host galaxy \citep{2015ApJ...802...89B,2017MNRAS.464.4545B,2017MNRAS.465.4348B}.

Beyond redshift 2, the obscured fraction was estimated to be larger than in the local universe but the behavior is still unclear. Some studies suggest that the obscured fraction is constant above redshift 2 \citep{2008A&A...490..905H,2014MNRAS.445.1430K,2016MNRAS.463..348V,2014MNRAS.445.3557V}. Also in contrast to the obscured fraction below redshift 2, it was suggested that the fraction of AGN with $\log N_{\rm H}\ (\rm cm^{-2})\geq 23$ at redshift 3 to 5 is independent of the X-ray luminosity \citep{2014MNRAS.445.3557V,2018MNRAS.473.2378V}.  Some studies found that the fraction of obscured AGN decreases with decreasing X-ray luminosity \citep{2015MNRAS.453.1946G}.

The large uncertainty in the fraction of obscured quasars above redshift 2 is partly due to the limited sample size associated with the limited survey area and depth of X-ray surveys. Survey depth is important for the detection of high redshift obscured AGN due to their X-ray faintness from obscuration and large distances. However, deep X-ray datasets are often limited to less than a few degrees of the sky. Furthermore, the number density of quasars beyond redshift 2 shows a strong decline with redshift thus a large survey area with sufficient depth is needed to construct a sizable sample of high redshift obscured quasars  \citep{2013ApJ...768..105M,2014MNRAS.445.1430K,2014MNRAS.445.3557V,2015MNRAS.453.1946G,2016MNRAS.463..348V,2018MNRAS.473.2378V}. 

While X-ray emission is a reliable tracer of accretion activity, it offers limited information of the AGN properties without distance estimates. Thus, X-ray sources must be matched with an optical/IR counterpart in order to determine their distance and properties, such as luminosity and column density of the nuclear obscuration. This means a large and deep multiwavelength dataset is needed to investigate the obscured fraction of quasars in the high redshift universe.

Recent development in large and deep X-ray surveys, which cover many legacy multiwavelength deep fields, has allowed the investigation of the obscured fraction at high redshift. Of particular interest is the XMM-Spitzer extragalactic representative volume survey (XMM-SERVS) in the XMM-LSS region \citep{2018MNRAS.478.2132C}, which is also covered by the deep optical imaging dataset from the Hyper Suprime-Cam Subaru strategic survey program (HSC-SSP; \citealt{2018PASJ...70S...4A}) and the deep $U$-band imaging data from the Canada France Hawaii Telescope large area $U$-band deep survey (CLAUDS; \citealt{2019MNRAS.489.5202S}) as well as previous legacy deep IR datasets.

In this study, the obscured fraction of luminous quasars above redshift 2 during the peak epoch of the quasar accretion growth was estimated by utilizing the unique wide and deep multiwavelength dataset within the XMM-SERVS region. The deep $U$-band image plays a crucial role to derive accurate photometric redshift for objects at $z>2$ with the Lyman break feature. Hereafter, obscured AGN refers to X-ray obscured AGN with $\log N_{\rm H}\ (\rm cm^{-2}) \geq 22$ unless stated otherwise.  We also investigate the correspondence between the X-ray obscuration and the restframe UV/optical spectral energy distribution (SED).  A galactic hydrogen column density of $3.57 \times 10^{20}\ \rm cm^{-2}$ in the survey area \citep{2018MNRAS.478.2132C} and a flat standard  $\rm \Lambda CDM$ cosmology with $H_0=70\ \rm km\ s^{-1}\ Mpc^{-1}$, $\Omega_M=0.3$, and $\Omega_\Lambda=0.7$ was assumed. Magnitudes are reported in the AB magnitude system.

\section{Data} \label{sec:data}

\subsection{XMM-SERVS} \label{sec:data:xservs}
\begin{figure*}[ht!]
    \centering
    \plotone{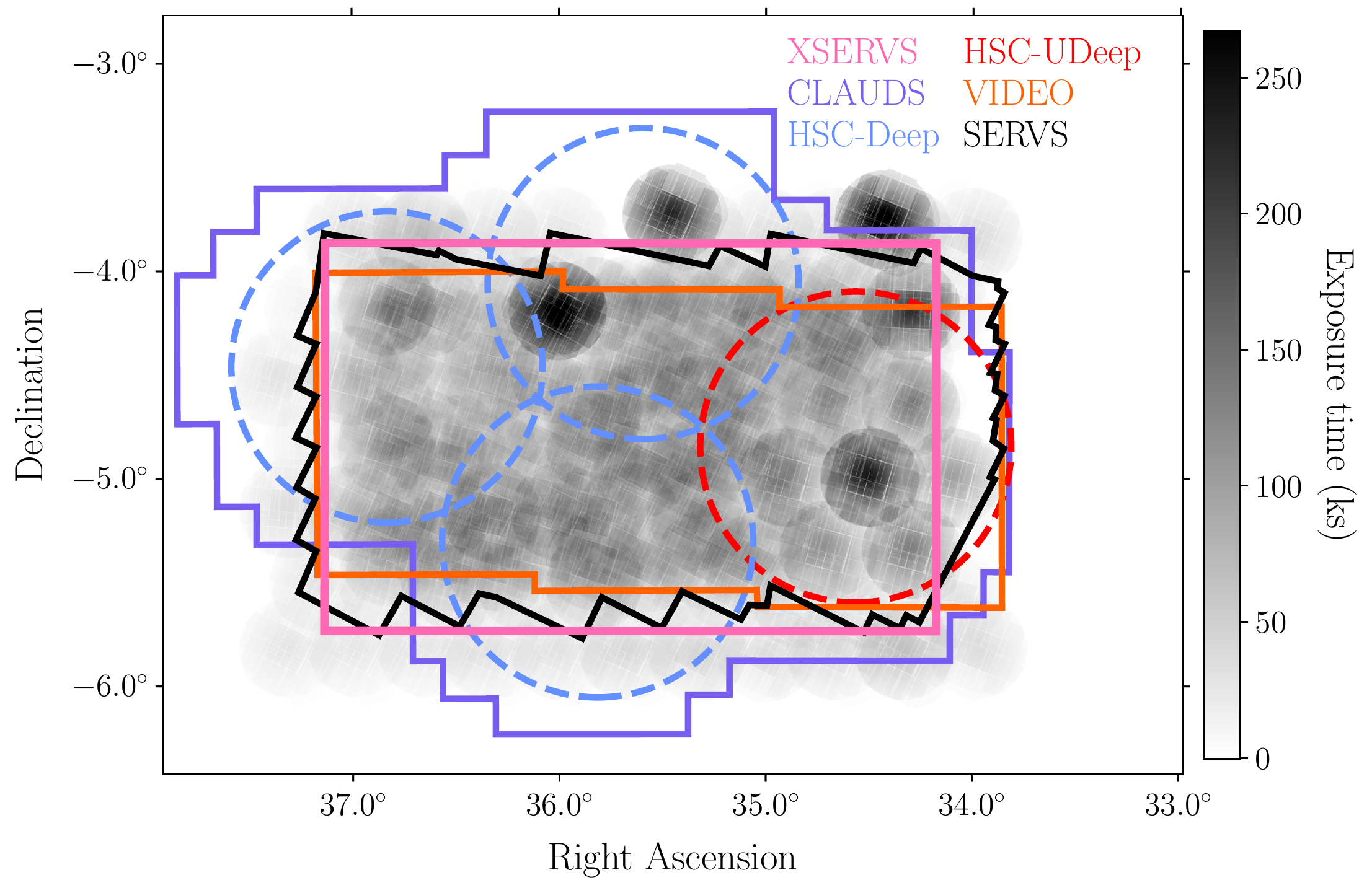}
    \caption{0.5-10 keV flare-filtered exposure map over the XMM-SERVS survey area \citep{2018MNRAS.478.2132C}. The catalog area of XMM-SERVS is shown in pink. The coverage of CLAUDS, HSC-Deep, HSC-UDeep, VIDEO, and SERVS are shown in purple solid, blue dashed, red dashed, orange solid, and black solid lines respectively. The color bar represents the commutative exposure time.}
    \label{fig:survey_coverage}
\end{figure*}

The XMM-SERVS X-ray point-source catalog in the XMM-LSS region \citep{2018MNRAS.478.2132C} was chosen as the primary selection of AGNs. The X-ray survey observations were performed by the {\it XMM-Newton} satellite over 5.3 square degrees of the XMM-LSS survey field. {\it XMM-Newton} has three detectors, MOS1, MOS2, and PN. The catalog contains the combined detection of 5242 X-ray point-sources using the three detectors in the 0.5-2 keV, 2-10 keV, and 0.5-10 keV bands. Figure \ref{fig:survey_coverage} shows the 0.5-10 keV flare-filtered exposure time within the survey area, which reaches $\sim50$ks per pointing continuously over the survey area. In some regions, deeper X-ray observations are available from other projects as listed in Table 2 of \citet{2018MNRAS.478.2132C}. The survey flux limits over 90\% of the total area are $ 1.7\times 10^{-15}, 1.3\times 10^{-14}$, and $6.5\times 10^{-15}\ \rm erg\ cm^{-2}\ s^{-1}$ in the 0.5-2 keV, 2-10 keV, and 0.5-10 keV bands, respectively. 
The flux from each detector in the catalog was derived using an energy conversion factor assuming a power-law continuum with photon index, $\Gamma=1.7$ and galactic absorption column density. The catalog provides the combined source flux calculated using the error weighted average of the flux estimated by all three detectors. In this paper, the count-rates were converted to the expected count-rates detected with the PN-detector (PN-equvalent count-rates) in the later discussions.  

\subsection{HSC-SSP} \label{sec:data:hsc}
The HSC-SSP is an optical imaging survey performed by the 8.2-meter Subaru telescope using the HSC imager \citep{2018PASJ...70S...1M}. HSC is a wide-field camera with a field of view of 1.5 square degrees. The camera is made up of 116 $\rm 2K\times4K$ fully-depleted back-illuminated CCDs (FDCCD; \citealt{2012SPIE.8453E..1XK}) mounted at the prime focus of the telescope. Among the HSC-SSP datasets with three different depths, imaging data of deep and ultra-deep layers in the XMM-LSS field are utilized from the S19A and S20A internal releases. Each HSC pointing is shown in Figure \ref{fig:survey_coverage} in red and blue dashed circles. S19A has better seeing in the HSC $r$-band than that in the S20A data release. The depth of the HSC deep layer is 27.4, 27.1, 26.9, 26.3, and 25.3 magnitudes in \textit{grizy} bands\citep{2022PASJ..tmp...12A}. The depth and typical seeing size are summarized in Table \ref{tab:survey_prop}. The depth was calculated from the median value of the 5-sigma point-source limiting magnitudes of all patches. Image data from HSC were reduced using the HSC pipeline \citep{2017ASPC..512..279J,2018PASJ...70S...5B,2019ASPC..523..521B,2019ApJ...873..111I}. Photometric and astrometric calibration was performed against the first data release of the Panoramic survey telescope and rapid response system (Pan-STARRS1; \citealt{2012ApJ...756..158S,2012ApJ...750...99T,2013ApJS..205...20M,2016arXiv161205560C,2020ApJS..251....6M}). The HSC images are warped onto predefined grids called ``Tracts" and ``Patchs". Each tract is approximately $\rm \rm 1.5\ deg \times 1.5 \ deg$ and divided into 9$\times$9 patches. 

\subsection{CLAUDS} \label{sec:data:clauds}
CLAUDS is a deep $U$-band imaging survey of the HSC deep layer performed by the 3.6m CFHT with Megacam \citep{2019MNRAS.489.5202S}. Megacam \citep{2003SPIE.4841...72B} is a wide field camera with 40 2048$\times$4612 pixel back-illuminated CCDs, and covers an area of 1.02 square degrees. 

The CLAUDS survey was performed using two $u$-band filters; $u^*$ and $U$. The two $u$-band filters are significantly different from each other. The new $U$-band filter has better transmission than the $u^*$-band filter and has no red-leak at $\rm 5000 \AA$, which is observed in the $u^*$-band filter. For XMM-LSS deep field, the CLAUDS survey was performed entirely using the $u^*$-band filter. The depth of the CLAUDS in the XMM-LSS reaches 26.6 magnitude (5 sigma detection in $2^{\prime\prime}$ diameter aperture), while in the ultradeep region, which corresponds to the Subaru XMM-Newton Deep Survey (SXDS; \citealt{2008ApJS..176....1F}) region, reaches 1 magnitude deeper. The survey properties are summarized in Table \ref{tab:survey_prop}.

Basic calibration and data reduction of CLAUDS Mega-Cam data were performed with the Elixir software \citep{2004PASP..116..449M} at CFHT. Elixir performs the basic data reduction before sending the data to the Canadian Astronomy Data Centre to be processed with MegaPipe \citep{2008PASP..120..212G}. MegaPipe performs astrometric calibration against Gaia astrometry while photometry was calibrated against SDSS $u$ band photometry, cross checked with synthetic $u$-band photometry produced using a combination of Pan-STARRS \citep{2020ApJS..251....6M} g-band and GALEX NUV photometry.

\subsection{VIDEO} \label{sec:data:video}
The VISTA deep extragalactic observations survey (VIDEO; \citealt{2013MNRAS.428.1281J}) is a deep near-infrared imaging survey performed by the 4.1 meter visible and infrared survey telescope for astronomy (VISTA) at Cerro Paranal with the VISTA InfraRed CAMera (VIRCAM; \citealt{2006SPIE.6269E..0XD}). VIRCAM consists of 16 $\rm 2K\times 2K$ Raytheon VIRGO HgCdTe detectors. 

VIDEO data-release  5 mosaic images of the XMM-SERVS field are provided in the ESO Phase 3 data archive.\footnote{http://eso.org/rm/publicAccess\#/dataReleases} In the XMM-SERVS field, the mosaic images in each band are separated into 3 smaller areas designated as XMM1, XMM2, and XMM3. Among them, XMM1 covers the SXDS.  The data were reduced at the Cambridge astronomical survey unit (CASU) using the VISTA data flow system (VDFS; \citealt{2004SPIE.5493..411I}). The astrometry and photometry of the survey were calibrated against the 2MASS point-source catalog \citep{2006AJ....131.1163S}. The final 5-sigma depths in a 2 arcsecond diameter aperture are 24.51, 24.44, 24.12, and 23.77 magnitudes in Y, J, H, and Ks bands, respectively. The survey properties are summarized in Table \ref{tab:survey_prop}.

\begin{deluxetable}{ccccc}[h]
\tablecaption{Properties of the deep optical and IR imaging datasets}
\tablewidth{0pt}
\tablehead{
    \colhead{Survey} & \colhead{Band} & \colhead{Area} & \colhead{Depth\tablenotemark{a}} &  \colhead{Seeing} \\
    \colhead{} & \colhead{} & \colhead{($\rm deg^{-2}$)} & \colhead{(mag)} &  \colhead{($^{\prime\prime}$)} 
}
\startdata
CLAUDS\tablenotemark{b} & $u^*$ & 8.78 & 26.60(27.60) & 0.92 \\
\hline
{  } & $g$ &  & 27.4 &  0.83(0.82) \\
HSC\tablenotemark{c} & $r$ &  & 27.1 &  0.58(0.72) \\
S19A & $i$ & 6.55 & 26.9 &  0.81(0.75) \\
{(S20A)} & $z$ &  & 26.3 &  0.82(0.79) \\
{  } & $y$ &  & 25.3 &  0.79(0.70) \\
\hline
{    } & $Y$ &  & 24.51 &  0.8 \\
VIDEO  & $J$ & 4.81 & 24.44 &  0.8 \\
{    } & $H$ &  & 23.12 &  0.8 \\
{    } & $K_s$ & & 23.77 & 0.8 \\
\hline
SERVS & $3.6\rm \mu \mathrm{m}$ & 5.62 & 23.20 &  1.6  \\
{    } & $4.5\rm \mu \mathrm{m}$ & 5.59 & 23.04 &  1.7 \\
\enddata
\footnotesize
\tablenotetext{a}{5-sigma detection limit in a $2^{\prime\prime}$ aperture.}
\tablenotetext{b}{Magnitude limit for the deep region, the number in the parenthesis represents that for the ultra-deep region.}
\tablenotetext{c}{Depth from \citealt{2022PASJ..tmp...12A}. The depth is defined as the 5-sigma limiting magnitude for pointsource objects \citep{2019PASJ...71..114A}. Seeing estimates are shown for S19A data release, the number in the parenthesis represents that of S20A data release and are from the internal data release data quality plots.}
\normalsize
\label{tab:survey_prop}
\end{deluxetable}

\subsection{SERVS} \label{sec:data:servs}

The SERVS \citep{2012PASP..124..714M} is a deep mid-infrared imaging survey performed by the Spitzer space telescope using the Infrared Array Camera (IRAC;\citealt{2004ApJS..154...10F}) during the post-cryogenic mission. Only the IRAC channel 1 ($\rm 3.6\ \mu \mathrm{m}$) and channel 2 ($\rm 4.5\ \mu \mathrm{m}$) are usable due to the high background in the other bands due to the outage of the cryogenic coolant.

SERVS covers 5 deep multiwavelength extragalactic fields (ELAIS-N1, ELAIS-S1, Lockman Hole, Chandra Deep Field South, and XMM-LSS), in total 18 square degrees. The mean integration time per pixel is approximately 1200s which is close to the confusion limit of the Spitzer IRAC data. The 5-sigma depths in 3.6 and 4.5 $\rm\mu \mathrm{m}$ bands are 1.9 and 2.2 $\rm \mu Jy$ in 3.8 arcsecond diameter aperture. They correspond to 5-sigma magnitudes of 23.20 and 23.04, respectively. The survey properties are summarized in Table \ref{tab:survey_prop}.

The mosaic and uncertainty images were retrieved from the NASA/IPAC Infrared Science Archive.\footnote{\citealt{https://doi.org/10.26131/irsa407}}. The data were processed at the Spitzer science center (SSC). The data reduction pipeline performs the standard image reduction and additional detector specific processing. The images were co-added and reprojected using MOPEX to a pixel scale of 0.6 arcsecond pixel\textsuperscript{-1}. Original photometric calibration of the IRAC data was performed using dedicated calibration observations. Crosschecks against the SWIRE survey\citep{2003PASP..115..897L} suggests that a correction factor of 1.02 is needed for the $3.6 \mu \mathrm{m}$ band but not for the $4.5 \mu \mathrm{m}$ band. We apply the correction during catalog construction. 

\subsection{Spectroscopic Redshifts} \label{sec:data:specz}

Similar to other multiwavelength survey fields, there are a large number of spectroscopic redshift measurements in the XMM-LSS region. Spectroscopic redshift measurements within the survey area were compiled from various spectroscopic surveys including those from the Sloan Digital Sky Survey data release 9 \& 16 (SDSS;\citealt{2020ApJS..249....3A,2012ApJS..203...21A}), VIMOS Public Extragalactic Redshift Survey (VIPERS; \citealt{2018A&A...609A..84S}), Galaxy and Mass Assembly (GAMA; \citealt{2015MNRAS.452.2087L}), VIMOS VLT Deep Survey (VVDS ; \citealt{2013A&A...559A..14L}), VANDELS \citep{2021A&A...647A.150G},  MOSFIRE Deep Evolution Field Survey (MOSDEF; \citealt{2015ApJS..218...15K}),  Ultradeep Survey\footnote{https://www.nottingham.ac.uk/astronomy/UDS/data/data.html} (UDS; \citealt{2013MNRAS.428.1088M,2013MNRAS.433..194B}), 3D-HST \citep{2012ApJS..200...13B,2016ApJS..225...27M}, and the SXDS multiwavelength catalog \citep{2015PASJ...67...82A}, as well as from individual studies in the SXDS and XMM-LSS regions (\citealt{2005ApJ...634..861Y,2007MNRAS.381.1369G,2008ApJS..176..301O, 2008ApJ...675.1076S,2008MNRAS.389..407S,2009MNRAS.395...11V,2010MNRAS.402.1580O,2012MNRAS.421.3060S,2013ApJ...771....7D,2013yCat..35570081M,2014MNRAS.437.3647Y,2016ApJ...819...24W,2016MNRAS.457..110M,2018PASJ...70S..10O}). In total,  294,536 secure spectroscopic redshift records associated with 238,403 unique galaxies and AGN were compiled. The majority of the spectroscopic redshift records are from the objects in the SDSS and VIPERS catalogs which have an $i$-band magnitude up to 22.5. However, the faintest magnitude of deep spectroscopic surveys such as VVDS-UDEEP, 3D-HST, and spectroscopic follow-up of X-ray sources in the SXDS from \citet{2015PASJ...67...82A} reaches to $i$-band magnitude of 24.75.

\section{Multi-band Photometry of The Optical Counterpart} \label{sec:datared}
\subsection{Process of Multi-band Photometry} \label{sec:datared:photometry}

In order to obtain the multiwavelength properties of the optical counterparts of the X-ray sources, this work uses deep imaging data in the 12 photometric filters from the datasets described above.  Proper treatment of the point-spread function (PSF) shape and size differences in between datasets is needed in order to obtain accurate colors for photometric redshift estimation and SED fitting. This is especially important for the SERVS mid-infrared dataset, which suffers from severe blending due to the larger PSF size than the other datasets. 

Prior-based PSF-convolved photometry was performed using T-PHOT \citep{2015A&A...582A..15M,2016A&A...595A..97M}. T-PHOT uses morphological information of an object in a high resolution image to measure its flux in images with low spatial resolution. The low resolution image needs to have the same world coordinate system (WCS) and the same or integers-times pixel-scale as the high resolution prior image. In this analysis, all data were resampled to the WCS defined in the HSC S19A internal dataset. The process of image alignment, background subtraction, and preparation of variance images are described in the following subsections.

\subsubsection{Image Alignment and Background Subtraction} \label{sec:datared:addimared}

At first, global background subtraction was applied to the calibrated HSC image in each patch. The background levels were determined from the mean pixel value of the image after applying a $\rm 3\sigma$ clip. After the background subtraction, an additional 200 blank pixels were padded to each side of the image.

CLAUDS data used in this work were aligned to the tract-patch definition as of HSC S16A dataset, a prior version to the HSC data currently used in this analysis. Therefore, there is a 1-2 pixel offset from the HSC images used in the current analysis, we match the astrometry to the HSC S19A dataset and apply an additional local background subtraction using SWarp \citep{2010ascl.soft10068B}. 

For the pipeline reduced VIDEO DR5 images of the XMM1, 2, and 3, the background of each area was subtracted using SExtractor \citep{1996A&AS..117..393B}. The images were then resampled into the same pixel scale as in the HSC images and combined together into a single mosaic using SWarp. Resampled variance images on the same pixel-scale were also produced using SWarp. These variance images are different from the weight images produced automatically by SWarp, which does not preserve the original variance. The variance images were created by converting the original weight images to variance and then passing them to SWarp as images to be combined. This method better preserves the original variance of the image than the weight images automatically generated by SWarp.  Cutouts in the same tract-patch as HSC images were created from the mosaics and resampled to the same WCS as in the HSC S19A images.

SERVS mid-infrared images are provided as a single mosaic with the RMS image. The mosaic image was resampled to the same WCS as in the HSC S19A images and the local background was subtracted using SWarp. The RMS image was converted to a variance image and resampled in the same manner applied to the VIDEO images. Cutouts in the same tract-patch and WCS as in the HSC S19A images were then produced from the mosaic images using SWarp. 

\subsubsection{PSF-Modeling \& Convolution Kernel Construction} \label{sec:datared:psfkern}

The convolution kernel is a 2-dimensional matrix that converts the PSF shape of the high-resolution image (HRI) to the PSF shape of the low-resolution image (LRI). It is constructed by deconvolving the LRI PSF with the HRI PSF. The PSFs were constructed locally in each patch in order to take into account the PSF variation over the survey area. However, PSF variation within each patch is ignored. The PSF of the HSC datasets were queried from the HSC database while the PSF models for CLAUDS, VIDEO, and SERVS were constructed directly from the images of stellar objects. 

At first, catalogs in each patch produced using SExtractor are matched with $Ks$ band sources in the 2MASS point-source catalog \citep{2006AJ....131.1163S}\footnote{\citealt{https://doi.org/10.26131/irsa2}}. For CLAUDS and VIDEO, 2MASS sources with $ 15 \leq K_s < 17$ were selected for the PSF modeling. Extended sources with $\rm CLASS\_STAR<0.9$ and $\rm ELLIPTICITY>0.2$ in the CLAUDS and VIDEO catalogs were removed. These extended sources were identified as point-sources in the 2MASS point-source catalog due to the low spatial resolution of 2MASS.

For SERVS, sources with $\rm 14 \leq K_s < 17$ were selected for the PSF modeling. Since the PSFs of Spitzer IRAC is asymmetric, the same criteria as the CLAUDS and VIDEO datasets cannot be used to remove extended sources. Extended sources were rejected by examining the FWHM distribution; sources whose FWHM exceeds $2\sigma$ scatter were removed.

After the removal of the extended sources, the individual images of stellar sources were re-centered, background  subtracted, and normalized to unity. The final PSF images were constructed using a median combination of each individual stellar source. Finally, the final PSF image is re-centered and normalized once again, and the images are padded to $55\times55$ pixels.

Local PSF images were constructed only for patches with more than 7 individual stellar sources. A median PSF image of each tract is used for patches with fewer than 7 individual sources. The HSC S19A $r$-band PSF images were adopted as the HRI PSF and Pypher \citep{2016A&A...596A..63B} was used to construct the convolution kernels of images in other bands.

\subsubsection{Pixel-Pixel Correlation Correction} \label{sec:datared:pxpx}
It is well known that pixel-pixel correlation occurs during image resampling with SWarp. Resampling smooths the image thus the variance in the image becomes artificially smaller than the original image. Therefore, it is necessary to correct the underestimation of the variance.

Fixed-aperture analysis similar to \citet{2012A&A...545A..23B} was performed on the background variance images to estimate the effects of pixel-pixel correlation. SExtractor was used to produce segmentation images of all CLAUDS and VIDEO images and 1000 fixed apertures of $2^{\prime\prime}$ radius were placed in regions with no source detections. The sky background was measured from the science images in the 1000 apertures while the photometric uncertainties were estimated from the weight images in the same aperture using PhotUtils \citep{larry_bradley_2020_4044744}.  A $4\sigma$ clip was used to remove outlier apertures that may have fallen on the image boundary or bad detector regions. The sky background variance ($\sigma_{sky}^2$) was estimated by fitting a Gaussian distribution to the distribution of the sky background values and compared the variance with the median of the variance image in the same aperture ($\sigma_{wei}^2$). If there is no pixel-pixel correlation, the two values are consistent with each other.

\begin{deluxetable}{cc}[!h]
\tablecaption{Pixel-pixel correlation correction factor for variance in the VIDEO dataset}
\label{tab:VIDEOcorrection}
\tablewidth{0pt}
\tablehead{
    \colhead{Band} & \colhead{Correction ($k=\sigma_{sky}^2/\sigma_{wei}^2$) } 
}
\startdata
VIDEO-Y & 16.78   \\
VIDEO-J & 11.88   \\
VIDEO-H & 11.56   \\
VIDEO-Ks & 8.32   \\
\enddata
\end{deluxetable}

The HSC pipeline has already performed the correction for pixel-pixel correlation and no resampling was done for the reduced image. For the SERVS dataset, thanks to the resampling process described in section \ref{sec:datared:addimared}, the SERVS variance data preserved correct variance and is consistent with the variance in the sky background.  On the other hand, The variances measured in CLAUDS and VIDEO was significantly smaller than the variance measured in the sky background.  The correction factor $k=\sigma_{sky}^2 / \sigma_{wei}^2$ was estimated in each patch of CLAUDS and VIDEO data.

\begin{figure}[ht!]
    \epsscale{1.1}
    \plotone{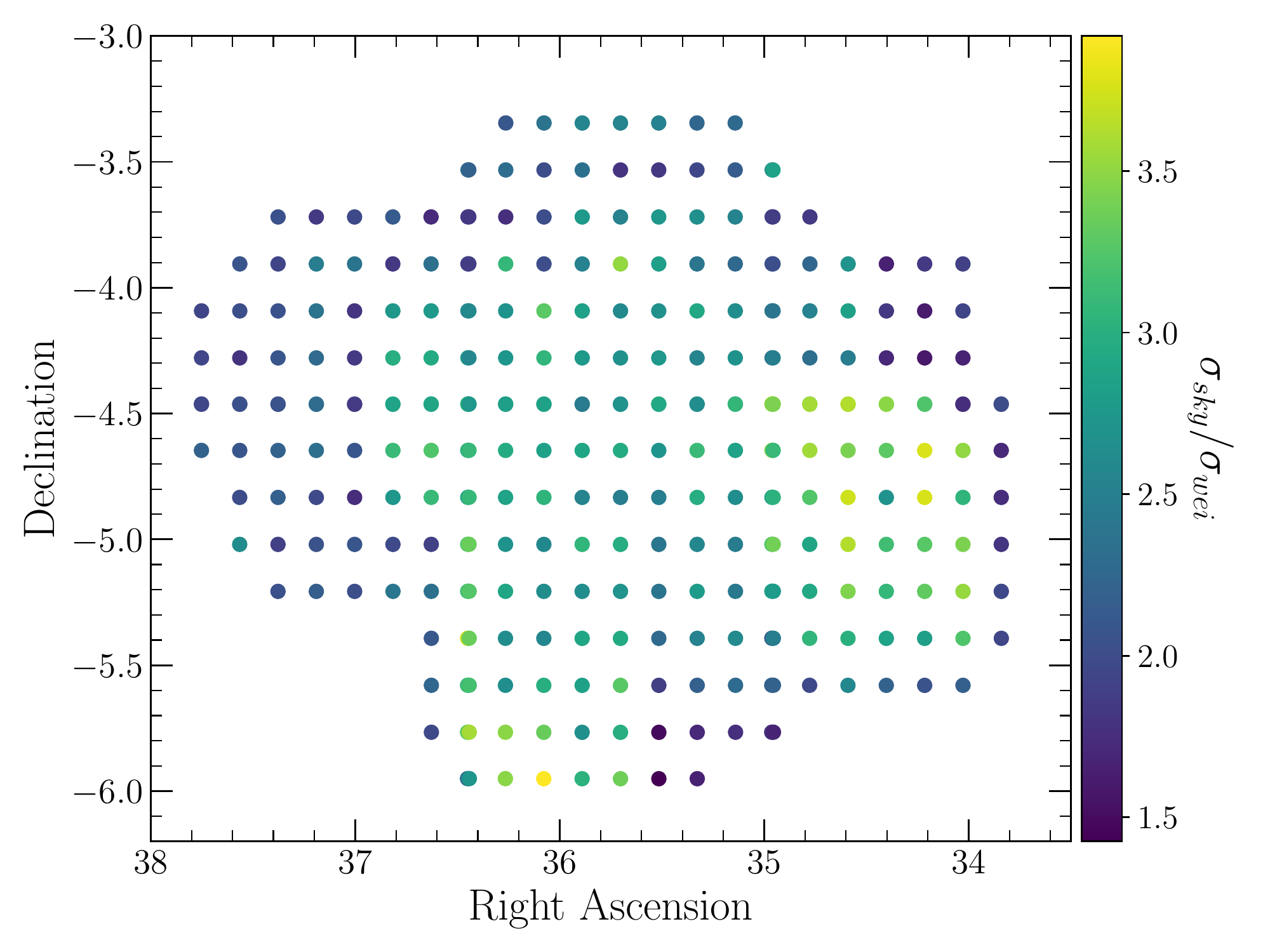}
    \caption{Distribution of the pixel-pixel correlation correction factor for variance in the CLAUDS data at each patch position.}
    \label{fig:clauds_corr}
\end{figure}

The correction factors for the VIDEO images are constant over the survey area of the VIDEO survey. A single correction factor determined from the median of the correction factors in all of the VIDEO images was adopted for simplicity and is summarized in Table~\ref{tab:VIDEOcorrection}. On the other hand, the correction factor of CLAUDS shows significant variation due to the variation in the survey depth.  The correction factor of each patch was applied individually and the median of all correction factors in each tract was used when the patch-level correction factor cannot be determined. Figure \ref{fig:clauds_corr} shows the distribution of the adopted correction factor over the CLAUDS survey area.

\subsection{Source Detection} \label{sec:datared:psfconv}
The primary source catalogs and segmentation images were constructed from HSC S19A $r$-band images using SExtractor. As summarized in Table \ref{tab:survey_prop}, HSC S19A $r$-band image has the highest resolution among the imaging datasets and is deep enough to detect a large fraction of objects in the other bands. Therefore, the $r$-band image is used as the high-resolution prior.

\begin{figure*}[t!]
    \plotone{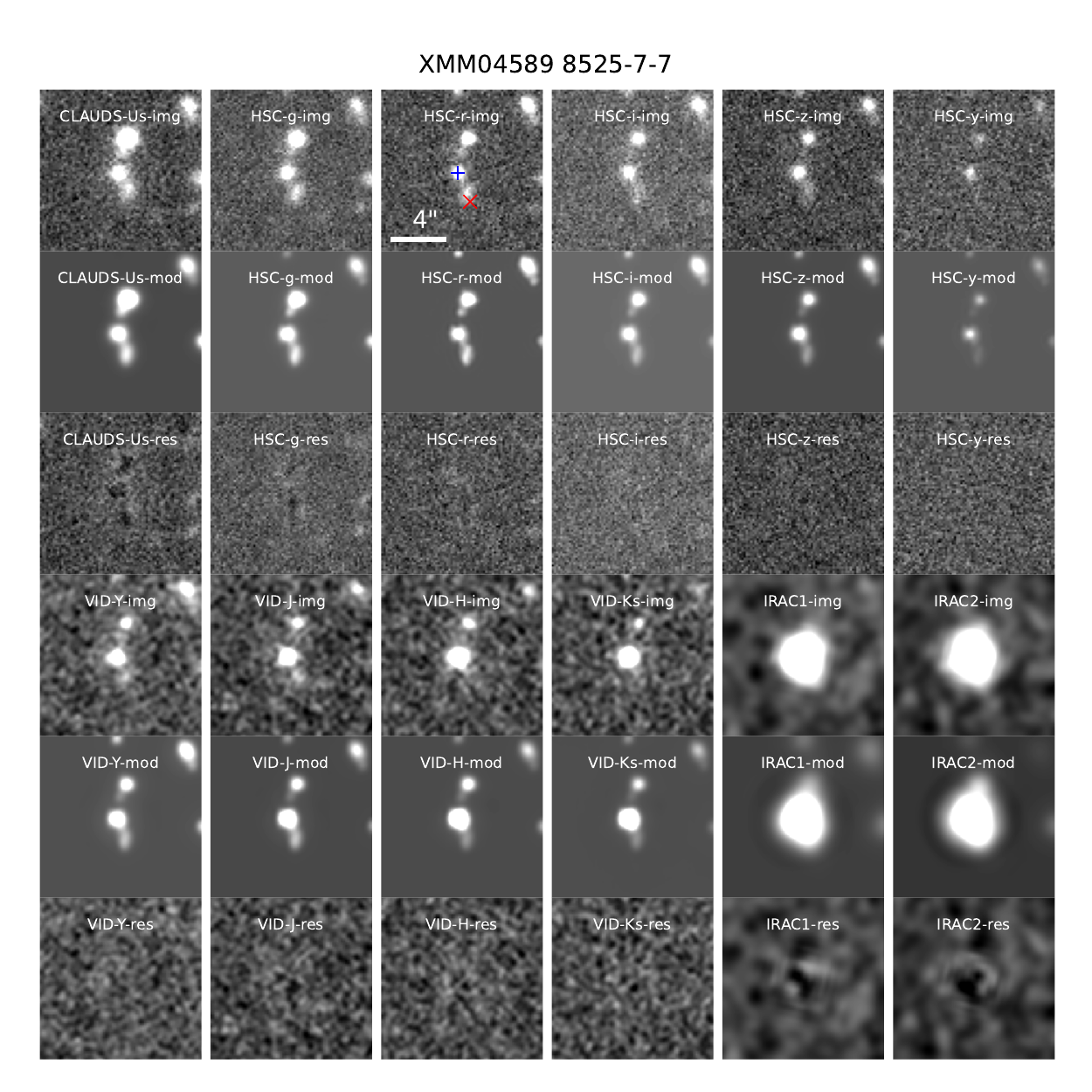}
    \caption{(First and fourth rows) cutout images of an optical-IR counterpart of an X-ray source in each photometric band. (Second and fifth rows) model images from the T-PHOT fitting process. (Third and sixth rows) residual images of the model fitting. The blue plus and red cross symbol mark the center of the optical-infrared counterpart and X-ray source center, respectively.}
    \label{fig:image_example}
\end{figure*}

The T-PHOT fitting process fails in some regions where bright stars are present in the image. This is likely due to the fact that the SExtractor deblending algorithm divides the star and halo into many individual sources. To solve this problem, sources within the HSC S19A $r$-band bright star masks \citep{2018PASJ...70S...7C} were removed.  T-PHOT was ran twice in each patch to take into account small sub-pixel astrometric offsets. The astrometric offsets were determined in the first run and applied automatically during the second run. The results include catalogs containing the fitting results, model images, residual images, and residual statistics.  

Figure \ref{fig:image_example} shows optical-IR images, model images, and residual images of an optical-infrared counterpart of an X-ray source. Several sources can be seen from $u^*$ to $Ks$ bands but are blended together in the 3.6 and 4.5 $\rm \mu \mathrm{m}$ images. The residual images in the 3.6 and 4.5 $\mu \mathrm{m}$ bands show systematic residuals. This is possibly due to the PSF asymmetry and its variation over the field of view of the IRAC datasets.

\subsection{Multi-band Photometric Catalog Creation} \label{sec:datared:catgen}

The T-PHOT catalogs were combined together and duplicate objects in overlapping patch regions were removed from the catalog. The photometric magnitudes and uncertainties in each band were calculated using the zero-points and zero-point uncertainties shown in Table \ref{tab:zp_list}. If the signal-to-noise ratio of the measurement is below $2\sigma$, then $2\sigma$ upper limits were adopted instead. 

Objects that fall into the HSC S20A bright star mask, bad detector regions affected by stray light, or detector defects, sources on the edge of the survey area, and sources with failed fitting results are flagged in the catalog. In addition, sources that are likely local galaxies and were broken up by the detection algorithm were also identified using region files created from the Hyper-LEDA catalog \citep{2014A&A...570A..13M}. Lastly, sources containing saturated pixels in the HSC images were flagged using the HSC mask images. The galactic reddening value for each object was retrieved from the IRAS reddening map\footnote{https://irsa.ipac.caltech.edu/applications/DUST/} of \citet{1998ApJ...500..525S}. The galactic dust attenuation in each band was calculated using the Galactic dust extinction law  \citep{1999PASP..111...63F}. 

\begin{deluxetable}{ccc}[!h]
\tablecaption{Photometric Zero-point}
\tablewidth{0pt}
\tablehead{
    \colhead{Band} & \colhead{Zero-point} & \colhead{Uncertainty} 
}
\startdata
CLAUDS-{$u^*$} & 30.0 &  0.035 \\
HSC-$g$ & 27.0 & 0.010  \\
HSC-$r$ & 27.0 & 0.010  \\
HSC-$i$ & 27.0 & 0.010  \\
HSC-$z$ & 27.0 & 0.011  \\
HSC-$y$ & 27.0 & 0.013  \\
VIDEO-YJHKs & 30.0 & 0.020 \\
SERVS 3.6 \& 4.5$\rm \mu \mathrm{m}$\tablenotemark{a} & 23.9 & 0.030 \\
\enddata
\tablenotetext{a}{Measurements were converted from $\rm MJy\ Sr^{-1}$ to $\rm \mu Jy\ pix^{-1}$}
\label{tab:zp_list}
\end{deluxetable}

\subsection{Survey Area} \label{sec:analy:area}

Because the multiwavelength photometry sample does not cover the entire sample of the X-ray point-sources in \citet{2018MNRAS.478.2132C}, we redefined the survey area of the X-ray sample based on the availability of the multiwavelength photometry considering the following conditions; 

\begin{enumerate}
    \item The optical-IR counterpart is in the HSC, VIDEO $H$-band, and SERVS-IRAC1 coverage.
    \item The optical-IR counterpart is not in any bright star masks of HSC nor in the bad regions of VIDEO $H$-band.
\end{enumerate}

\begin{figure*}[ht!]
    \centering
    \epsscale{1.}
    \plotone{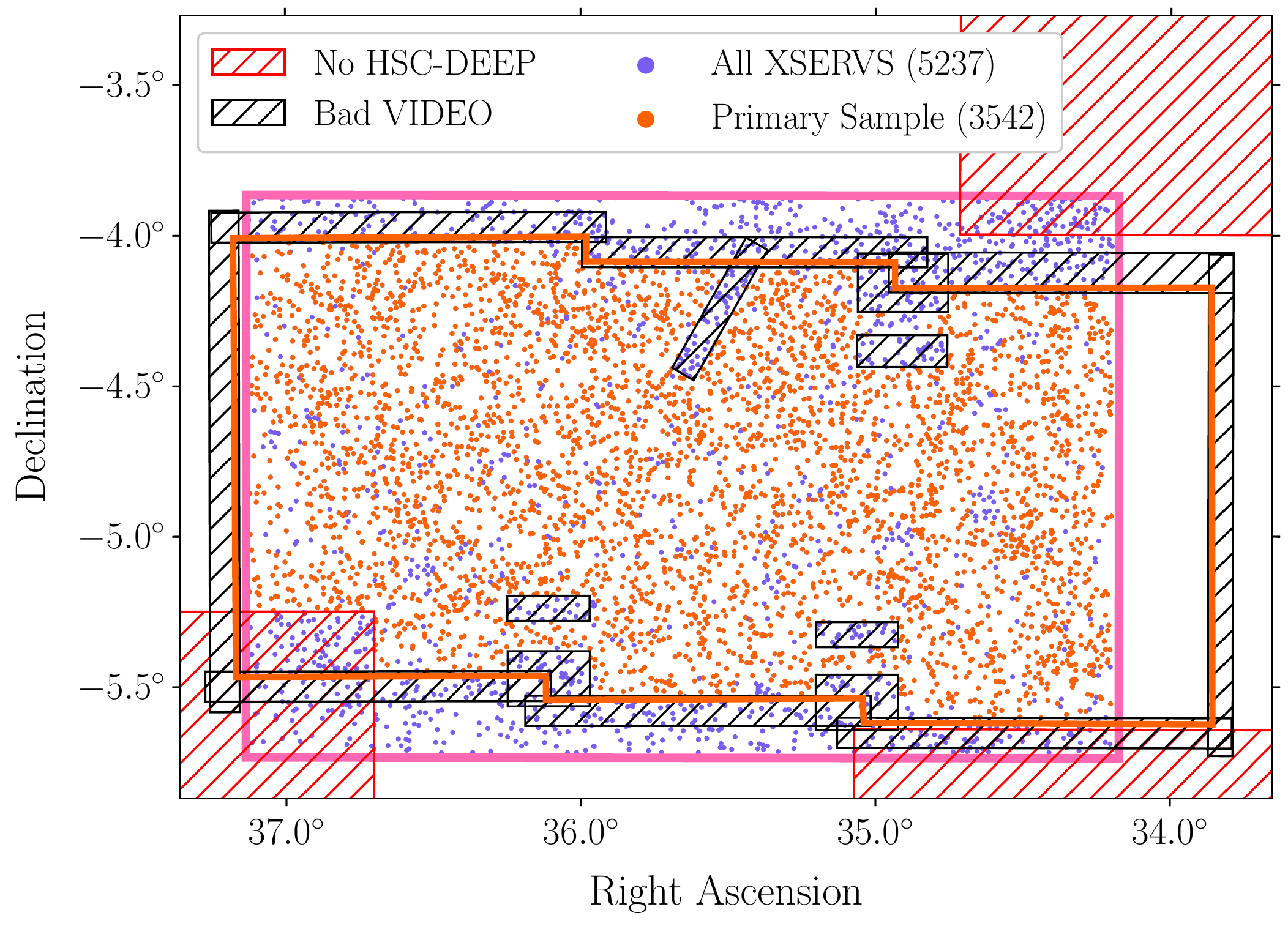}
    \caption{Redefined survey area compared to the XMM-SERVS X-ray catalog area (\citealt{2018MNRAS.478.2132C}, pink box). X-ray sources in the catalog are shown in purple, while the X-ray sources which satisfy the redefined survey area criterion are shown in orange. Regions with no HSC-Deep data are shown in red hatched regions. Regions affected by stray light or near the edge in VIDEO $H$-band images are shown in black hatched areas. Purple points within the multi-band coverage are X-ray sources around bright optical stars.}
    \label{fig:redef_survey}
\end{figure*}

The first condition was imposed to maximize the coverage of multiwavelength photometry, while the second condition was imposed to remove regions where the optical-IR images were affected by image artifacts such as stray light, bright star halos, and edges of the images. Figure \ref{fig:redef_survey} shows the distribution of the X-ray sources that meet the above criteria shown as orange symbols. Out of the 5237 XMM-SERVS X-ray sources with an optical-IR counterpart, 3542 X-ray sources are selected as the primary sample for the statistical discussion.

The total area within the redefined area was estimated using a Monte Carlo simulation by randomly distributing 100,000 mock data-points within the original survey area presented in \citet{2018MNRAS.478.2132C} and calculating the fraction of data-points that satisfy the criteria. The estimated survey area for statistical analysis is 3.52 square degrees. The area curve in the 0.5-2 keV, 2-10 keV, and 0.5-10 keV bands was calculated by normalizing the maximum area of the area curve presented in \citet{2018MNRAS.478.2132C} to be 3.52 square degrees.

\begin{figure*}[ht!]
    \centering
    \epsscale{1.}
    \plotone{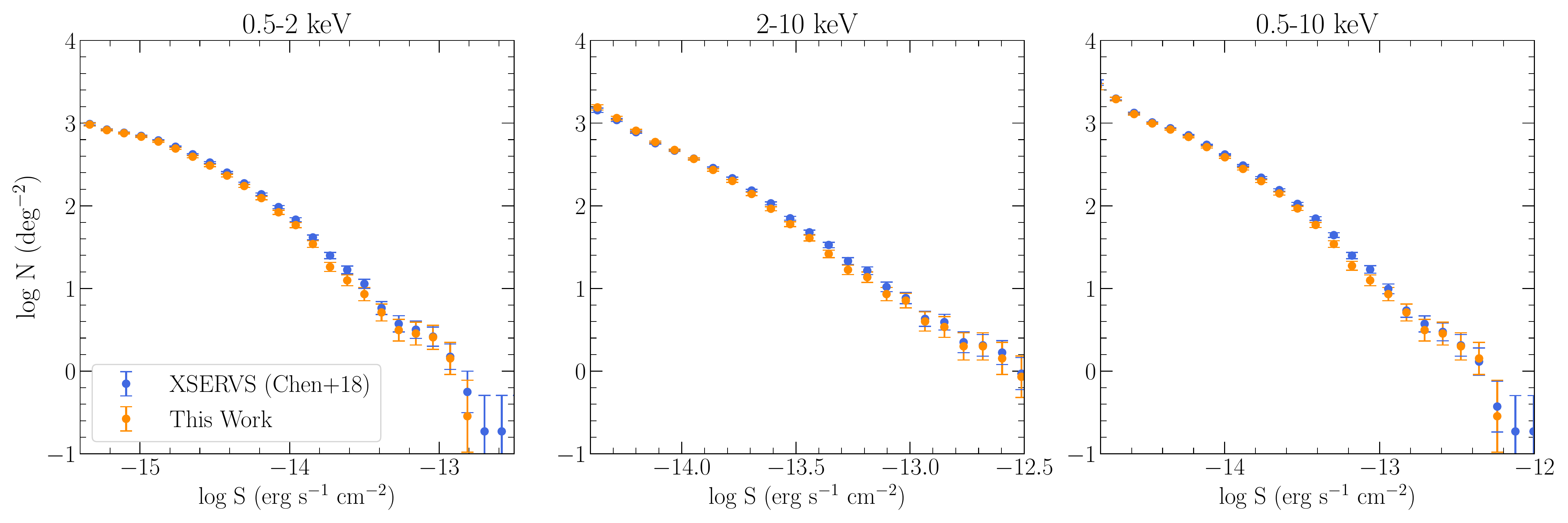}
    \caption{$\log N-\log S$ of the primary sample with the redefined survey area (orange) compared with that of the original sample of the  XMM-SERVS (blue) in the 0.5-2 keV, 2-10 keV, and 0.5-10 keV bands.}
    \label{fig:logNlogS}
\end{figure*}

In order to check the updated area curve, the logN-logS of the primary sample based on the redefined survey area is compared to that of the entire XMM-SERVS sample. Figure \ref{fig:logNlogS} shows the $\log N-\log S$ relation in the 0.5-2 keV, 2-10 keV, and 0.5-10 keV bands for the primary sample (orange) and the original XMM-SERVS (blue). The $\log N-\log S$ based on the redefined survey area is consistent with the $\log N-\log S$ of the original survey area within the Poisson uncertainty. We conclude that the normalized survey area reproduces the survey area of the primary sample well.

\subsection{Cross Matching with the X-ray Catalog}

In order to select the optical counterpart of each X-ray source, at first, the identification in \citet{2018MNRAS.478.2132C} was adopted. The original optical counterparts of 3180 X-ray sources have a corresponding object in the PSF-convolved photometric catalog within a $2^{\prime\prime}$ radius. In \citet{2018MNRAS.478.2132C}, the majority of X-ray sources (2762 sources) were matched with counterparts in the SERVS catalog. The SERVS-matched counterparts sometimes contain multiple counterparts in the photometric catalog due to the large PSF size of IRAC. For each SERVS-matched source, we examined the number of $r$-band detected neighbors within a $2^{\prime\prime}$ radius of the counterpart,  342 sources have multiple optical counterparts suggesting they are blended. The brightest source in the SERVS $3.6\mu \mathrm{m}$ band was chosen as the counterpart of the X-ray source among the blended sources. This modification changed the optical counterpart of 23 of the blended sources.

The remaining 362 sources with no corresponding counterpart within the PSF-convolved catalog show none or faint objects in the HSC S19A $r$-band but show a significant detection in the NIR bands. In order to recover the optically-faint counterparts, we produced PSF-matched cutouts by matching the PSF to that of VIDEO $H$-band using the PSF described in Section \ref{sec:datared:psfkern}. Aperture photometry using $2^{\prime\prime}$ diameter aperture was performed using SEextractor in dual-image mode on the optical and near-infrared images with the VIDEO $H$-band image as the detection image. The $2^{\prime\prime}$ diameter aperture contains $\sim 80\%$ of the PSF flux. Aperture correction factors were calculated from the VIDEO H-band growth curves. The factor was calculated for each patch and applied to the patch individually in order to account for the PSF variation. For the mid-infrared datasets, prior-based PSF convolved photometry was performed using the VIDEO $H$-band images as the high-resolution images. Photometry for 282 of the 362 HSC S19A $r$-band non-detected sources were successfully obtained. 

The images of the remaining 80 sources show neither HSC S19A $r$-band nor VIDEO $H$-band source but some show faint HSC S19A $i$-band objects close to the detection limit. We do not attempt to recover these sources because the optical identification with such faint sources is uncertain. The photometry of the 282 VIDEO $H$-band detected sources and 3180 $r$-band detected sources were combined together into a single catalog. In summary, the multi-wavelength photometry is obtained for $97.7\%$, 3462 out of the 3542 primary X-ray sources.  The catalog description of the primary X-ray AGN sample and multiwavelength photometry in the HSC-DEEP XMM-LSS region is provided in Appendix \ref{appendix}.

\section{Analysis}

\subsection{Photometric Redshift} \label{sec:analy:photoz}

Out of the 3462 primary X-ray sources, 1321 sources have prior spectroscopic redshift measurements. For the remaining X-ray sources, we calculated the photometric redshift using the photometric redshift code LePhare \citep{1999MNRAS.310..540A,2006A&A...457..841I}. LePhare estimates the photometric redshift by minimizing the $\chi^2$ between the observed and model photometry which was derived from template SEDs. For X-ray detected sources, empirical templates of galaxies, local AGN, as well as composite AGN templates constructed from AGN and galaxy templates in \citet{2011ApJ...742...61S} were used. For X-ray non-detected sources, galaxy models from \citet{2009ApJ...690.1236I} were used. The model magnitudes were calculated between redshift 0 and 6 with steps of 0.05. We use both SMC and Calzetti extinction laws \citep{1984A&A...132..389P,1994ApJ...429..582C} and fit the reddening as a free parameter with E(B-V) of 0, 0.025, 0.005, 0.075, 0.01, 0.02, 0.03, 0.04, 0.05, 0.06, 0.07, 0.08, 0.09, 0.1, 0.125, 0.15, 0.175, 0.2, 0.25, 0.3, 0.35, 0.4, 0.45, 0.5, 0.55 and 0.6 mag. No additional emission lines were considered for the AGN models but are added to the galaxy models. LePhare has the capability to estimate systematic zero-point shifts and apply them to the photometry to reduce the deviation from spectroscopic redshifts. The systematic shifts are derived using the spectroscopic redshift sample. The zero-point correction determined from X-ray non-detected sources was adopted to the X-ray detected sources and is shown in Table \ref{tab:zp_shift}. In both cases, an additional uncertainty (ERR\_SCALE) of 0.05 mag was also adopted in order to take into account any additional systematic uncertainty associated with the photometry.

\begin{deluxetable}{cccc}[!h] \label{tab:zp_shift}
\tablecaption{LePhare Systematic Zero-point Shifts}
\tablewidth{0pt}
\tablehead{
    \colhead{Band} & \colhead{Shift} & \colhead{Band} & \colhead{Shift} 
}
\startdata
CLAUDS-{$u^*$} & 0.156 & VIDEO-Y & 0.002 \\
HSC-$g$ & -0.027 & VIDEO-J & 0.049 \\ 
HSC-$r$ & -0.037 & VIDEO-H & 0.070 \\
HSC-$i$ & -0.0254& VIDEO-Ks & -0.021 \\
HSC-$z$ & -0.022 & SERVS-3.6$\mu$m & -0.005 \\
HSC-$y$ & -0.044 & SERVS-4.5$\mu$m & 0.003 \\ 
\enddata
\end{deluxetable}

\begin{figure*}
    \centering
    \epsscale{1}
    \plotone{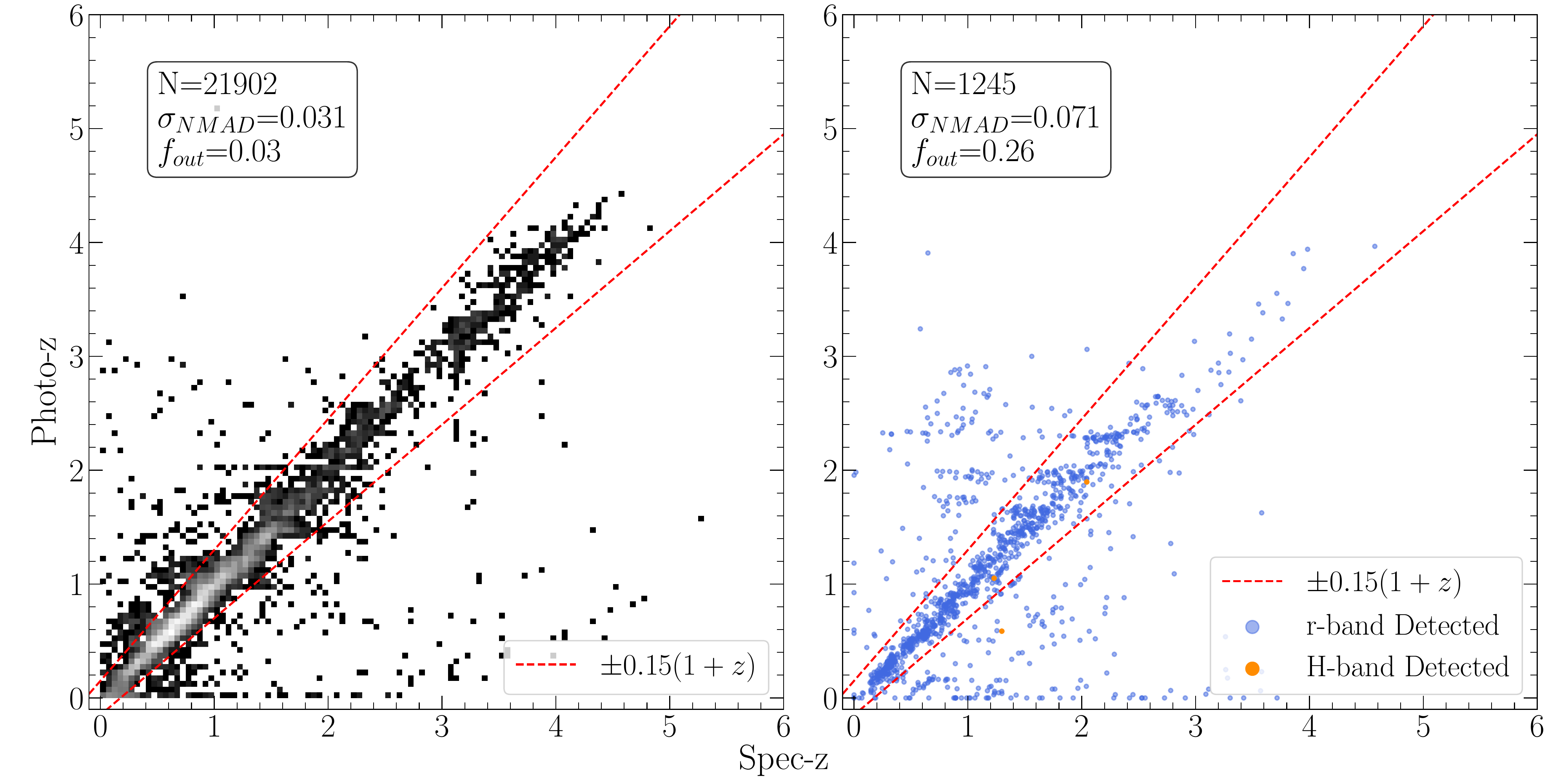}
    \caption{Photometric redshift compared with spectroscopic redshift measurements of galaxies without X-ray detection (left) and X-ray detected galaxies (right). The photometric redshift estimation performance statistics are shown at the top left.}
    \label{fig:photozcompare}
\end{figure*}

 Photometric redshift performance was evaluated based on the median absolute deviation ($\sigma_{\rm NMAD}$: \citet{1983ured.book.....H}) defined as $$\sigma_{\rm NMAD}=1.48\times {\rm Median}\bigg( \frac{|z_{\rm phz}-z_{\rm spec}|}{1+z_{\rm spec}}\bigg)$$ where $z_{\rm phz}$ and $z_{\rm spec}$ are the photometric redshift and spectroscopic redshift, respectively. The outlier fraction ($f_{\rm out}$) is the fraction of objects whose normalized absolution deviation $(z_{\rm phot}-z_{\rm spec})/(1+z_{\rm spec})$ is larger than 0.15 in the spectroscopic redshift sample. A threshold of 0.15 was adopted following studies in the COSMOS field \citep{2009ApJ...690.1236I,2016ApJS..224...24L}.
The photometric redshift performance was evaluated using sources that do not lie within the stellar locus in the $g-z$ against $z-3.6\mu \mathrm{m}$ plane $$(g-z)-0.5937(z-[3.6])<-1.7$$ where $g,z$ and [3.6] are the $g$, $z$, and 3.6$\mu \mathrm{m}$ band magnitudes. Galaxy templates were used to evaluate the photometric redshift performance of the X-ray non-detected sources. Similarly, AGN or galaxy templates were used to evaluate the photometric redshift performance of the X-ray detected source. Figure \ref{fig:photozcompare} shows the comparison between photo-z and spec-z of X-ray non-detected (left) and X-ray detected sources (right).

A photometric redshift scatter ($\sigma_{\rm NMAD}$) of 0.07 and an outlier fraction ($f_{\rm out}$) of 26\% was achieved for X-ray detected sources with 4 catastrophic failures where the photometric redshift cannot be determined. For comparison, a photometric redshift scatter $\sigma_{\rm NMAD}$ of 0.031 and outlier fraction $f_{\rm out}$ of 3\% was achieved for the X-ray non-detected galaxies.  The $\sigma_{\rm NMAD}$ and $f_{\rm out}$ for X-ray detected sources are worse than X-ray non-detected sources. Photometric redshifts for AGN can be difficult to accurately determine due to the flat featureless UV-optical continuum of unobscured AGN. 

\begin{figure}[ht!]
    \centering
    \epsscale{1}
    \plotone{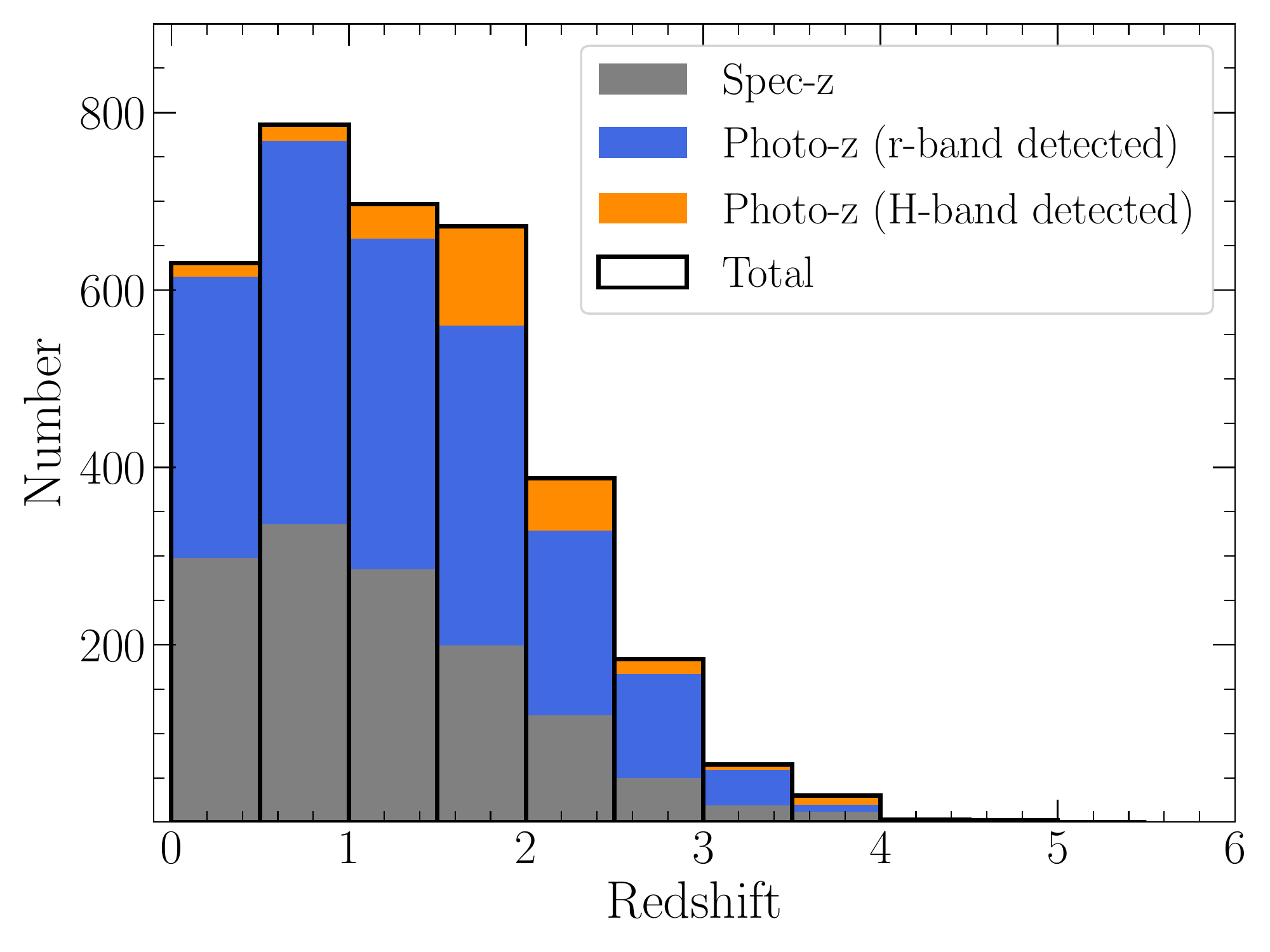}
    \caption{Redshift distribution of the X-ray sources shown as a stacked histogram. Spectroscopic redshift, photometric redshift from $r$-band detected source, and photometric redshift from $H$-band selected sources are shown in gray, blue, and orange  from bottom to top, respectively.}
    \label{fig:zdist}
\end{figure}

Figure \ref{fig:zdist} shows the distribution of the spectroscopic and photometric redshift of the primary X-ray sources as a stacked histogram. Above redshift 2, most of the redshift estimates are from the photometric redshift.  In summary, 3458 AGN of the primary sample have a spectroscopic redshift or 12-band photometric redshift, thus we achieved a total 99.8\% redshift completeness thanks to the deep multiwavelength dataset. 

\subsection{Hydrogen Column Density Estimation} \label{sec:analy:nhcalc}
The hydrogen column density ($\log N_{\rm H}$) associated with the nuclear X-ray emission of the primary X-ray source was calculated using the X-ray hardness ratio (HR) and best redshift estimation assuming an intrinsic AGN X-ray spectrum. The HR is defined as $(H-S)/(H+S)$ where $S$ is the 0.5-2 keV band count-rate and $H$ is the 2-10 keV band count-rate. In the XMM-SERVS catalog, the flux of each source is derived by combining measurements from multiple detectors with weighting. The PN-equivalent count-rate was calculated  by dividing the flux with the energy conversion factor (ECF) of PN used in \citet{2018MNRAS.478.2132C}. As a check, the PN-equivalent count-rate was compared with the reported count-rate of sources detected with the PN detector in  \citet{2018MNRAS.478.2132C}. The two count-rates are consistent with each other for the 2-10 keV band but show a systematic offset in the 0.5-2 keV band. A correction factor of 1.12 was applied to the 0.5-2 keV band to make the PN-equivalent count-rate consistent with that measured with the PN detector.

\begin{figure*}[ht!]
    \centering
    \epsscale{1.1}
    \plotone{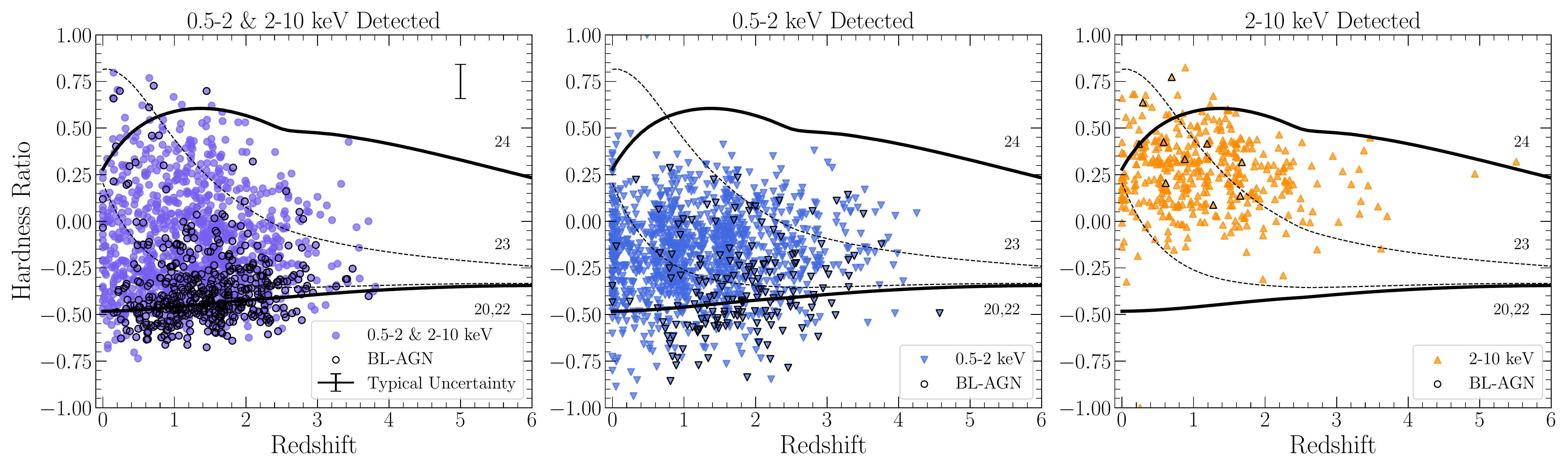}
    \caption{X-ray HR of AGN detected in both of the 0.5-2 keV and 2-10 keV bands (left, purple round symbols), AGN detected only in 0.5-2 keV (middle, blue downward triangle), and AGN detected only in 2-10 keV (right, orange upward triangle). BL-AGN are marked with black-rimmed symbols. The typical uncertainty is shown with an error bar at the upper right corner. The predicted HR of AGN with $\log N_{\rm H}\ (\rm cm^{-2}) =20$ and  $24$ are shown with solid lines while $\log N_{\rm H}\ (\rm cm^{-2}) =22$ and $23$ are shown with dashed lines.}
    \label{fig:z_hr_dist}
\end{figure*}

A phenomenological AGN model constructed from a linear combination of a cutoff power-law, pexrav reflection component, and scattered AGN continuum was assumed as the AGN X-ray spectra. 
\begin{equation*}
\begin{split} 
\rm Model:=& \rm tbabs\cdot(zphabs\cdot cabs \cdot zcutoffpl \\  & \rm + pexrav+ constant\cdot zcutoffpl) 
\end{split}
\end{equation*} 
where tbabs is the galactic X-ray extinction and zphabs is the absorption associated with the AGN, cabs is the additional Compton scattering, and constant is the fraction of the scattered AGN continuum. The parameters of the phenomenological model were set according to the best-fitted values from \citet{2017ApJS..233...17R} based on AGN in the local universe assuming a photon index of $\Gamma=1.8$. For the reflection component, the reflection strength was set to 1.0 and the inclination angle was set to 30 degrees. The fraction of the scattered AGN continuum was set to 1\% of the transmitted AGN continuum. The exponential cut-off was set to 381 keV for all components. The elemental abundances were set to solar abundance of \citet{1989GeCoA..53..197A}.

The HR model grid was calculated in the redshift range between 0.001 to 6 and $\log N_{\rm H}\ \rm(cm^{-2})$ grid between 20-26 using XSPEC \citep{1996ASPC..101...17A} assuming an on-axis response matrix file (RMF) and ancillary matrix files (ARF) of the {\it XMM-Newton} PN detector.\footnote{https://www.cosmos.esa.int/web/xmm-newton/epic-response-files} Figure \ref{fig:z_hr_dist} shows the comparison between HR of the primary sample and the HR of the model grid. Spectroscopically-identified type-1 broad-line AGN (BL-AGN) are marked with open circles, their HR shows a mild increasing trend with increasing redshift. Since most of the BL-AGN should have low column density ($\log N_{\rm H}\ \rm (cm^{-2}) < 22$), the increasing X-ray hardness is unlikely driven by increasing obscuration. The increasing X-ray hardness can be explained by the Compton reflection component at restframe 10-30 keV shifting into the observed hard 2-10 keV band at high redshift. It should be noted that broad-line classification is available only for a limited sample.

The column density was estimated only for AGN above redshift two and only between $\log N_{\rm H}\ (\rm cm^{-2})=20-24$ where the HR can distinguish the obscuration of CTN-AGN. The redshift dependence of the HR indicates that obscured AGNs with column density less than $\log N_{\rm H}\ (\rm cm^{-2})<22$ can hardly be distinguished. The column density can only be constrained for sources detected in both of the 0.5-2 keV and 2-10 keV bands. Only upper or lower limits on the column density can be derived for X-ray sources detected only in a single band. It should be reminded that the lower limit of the column density for sources that are detected only in the 2-10 keV is mostly larger than  $\log N_{\rm H}\ (\rm cm^{-2})>23$ above $z>2$. Therefore, sources detected only in the 2-10 keV band were assumed to have $\log N_{\rm H}\ (\rm cm^{-2})>23$ in the following discussion. Furthermore, some AGN show a smaller HR than that of an AGN model with $\rm \Gamma=1.8$. These sources likely have a softer AGN X-ray spectrum with a larger photon index ($\Gamma >1.8$) and no obscuration. We assign a column density of $\log N_{\rm H}\ (\rm cm^{-2})=20$ to them. 

\subsection{High Redshift AGN Sample Selection} \label{sec:analy:sample}
In order to examine the obscured fraction of luminous AGNs above redshift 2, 673 AGN between redshift 2-5 were selected based on the best available redshift (either spec-z or photo-z) as the high redshift AGN sample. Of the 672 AGN, 251 AGN were detected in both of the 0.5-2 keV and 2-10 keV bands, while 275 and 53 were detected only in the 0.5-2 keV and 2-10 keV bands, respectively. Finally, 93 AGN were detected only in 0.5-10 keV band. Within the 672 AGN, 203(30\%) have spectroscopic redshift and more than half of the spectroscopic sample are BL-AGN(71\%).

The intrinsic X-ray 2-10 keV luminosity was derived using 

$$L_X ={\rm ext}(z,N_{\rm H})  k(z)  4\pi D_L^2(z)  f_X$$

by assuming the best available redshift and $\log N_{\rm H}$, where $D_L$ is the luminosity distance, $f_X$ is the observed X-ray flux in 2-10 keV band, $k(z)$ is the k-correction term, and ${\rm ext}(z, N_{\rm H})$ is the extinction correction in the observed frame. The observed fluxes were calculated from the PN-equivalent count-rate in 2-10 keV band assuming an ECF of $1.26\times 10^{11}\ \rm counts^{-1}\ /\ erg\ s^{-1}cm^{-2}$.\footnote{For 0.5-2 keV and 0.5-10 keV bands, the ECF are $6.84\times 10^{11}$ and $3.36\times 10^{11} \ \rm counts^{-1}/erg\ s^{-1}cm^{-2}$, respectively} The ECF and extinction correction was calculated using XSPEC by assuming the phenomenological AGN model presented in Section \ref{sec:analy:nhcalc}. The ECF was from the model without absorption while the extinction correction was the ratio between the model without absorption to that with  $\log N_{\rm H}=20-24$.

\begin{figure*}[ht!]
    \centering
    \epsscale{1.1}
    \plotone{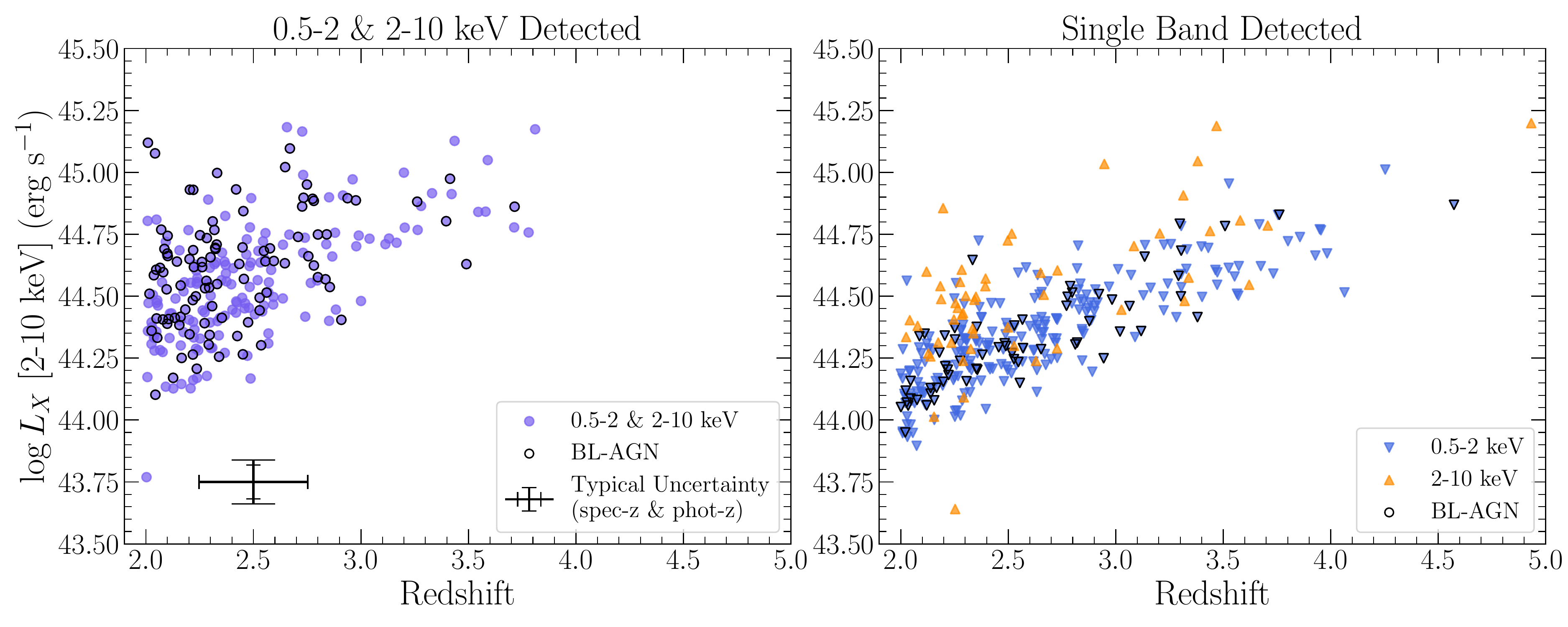}
    \caption{(Left) The intrinsic 2-10 keV luminosity of high redshift AGN detected in both 0.5-2 keV and 2-10 keV as a function of redshift (purple). BL-AGN are shown  with black open circles. The typical uncertainty for spec-z and photo-z samples is shown with the errorbar with sort and long caps, respectively. (Right) Same as left but upper limits and lower limits for AGN detected only  0.5-2 keV band (blue downward triangles) or 2-10 keV band (orange upward triangles).}
    \label{fig:z_lx_dist}
\end{figure*}

\begin{figure*}[ht!]
    \centering
    \epsscale{1.1}
    \plotone{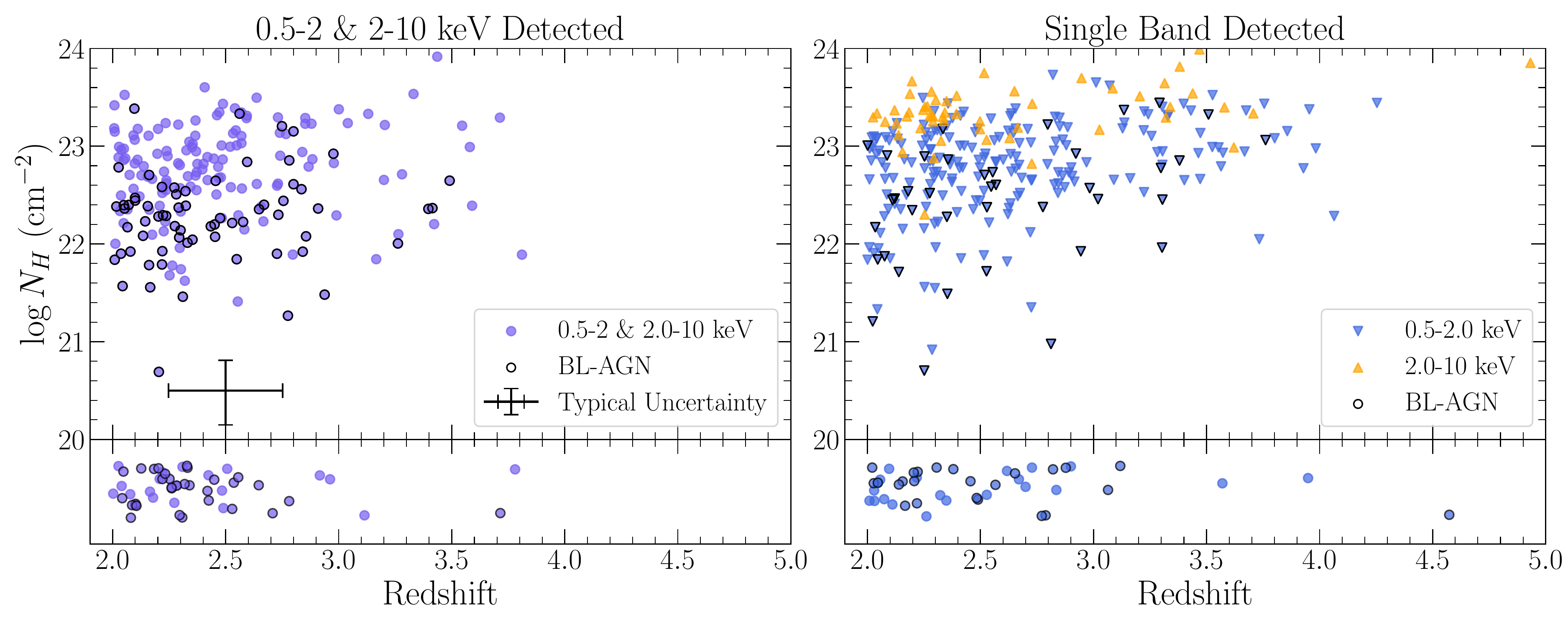}
    \caption{(Left) $\log N_{\rm H}$ of high redshift AGN detected in both 0.5-2 keV and 2-10 keV as a function of redshift (purple). BL-AGN are shown  with black open circles. The bottom panel shows the redshift distribution of AGN with HR softer than the minimum HR calculated by XSPEC are set to have $\log N_{\rm H}\ \rm (cm^{-2})=20$, the data-points in this panel are randomly shifted in y-axis for clarity. The typical uncertainty is shown with the errorbar at the bottom left corner. (Right) Same as left but upper limits and lower limits for AGN detected only  0.5-2 keV band (blue downward triangles) or 2-10 keV band (orange upward triangles).}
    \label{fig:z_nh_dist}
\end{figure*}

Figure \ref{fig:z_lx_dist} and \ref{fig:z_nh_dist} show the 2-10 keV luminosity and $\log N_{\rm H}$ of the high redshift AGN. BL-AGN are marked with a black open circle.  Most of the AGN detected in both of the 0.5-2 keV and 2-10 keV bands are those with $\log N_{\rm H}\ (\rm cm^{-2})<23.5$ and posses quasar level luminosity ($\log L_X\ (\rm erg\ s^{-1})>44.5$).  For AGN detected only in the 0.5-2 keV band, only upper limits can be placed on $\log N_{\rm H}$ and  $\log L_X$ and the upper limits show a large scatter. On the other hand, the lower limits of  $\log N_{\rm H}$ and $\log L_X$ are derived for AGN detected only in the 2-10 keV band. Most of them are located above $\log N_{\rm H}>23$, which implies that they are heavily obscured AGN.

\subsection{LDDE model \& Absorption Function} \label{sec:analy:lddemod}

In order to model the intrinsic number of AGN at $z>2$, the functional form of the luminosity and absorption functions were assumed following \citet{2014ApJ...786..104U} because the size and luminosity coverage of the current sample are not large enough to determine the overall shape of the luminosity function. The hard X-ray AGN luminosity function describes the number density of CTN-AGN with $\log N_{\rm H}\ (\rm cm^{-2})=20-24$. It is expressed as an evolving double power-law following the luminosity depended density evolution (LDDE) model. The hard X-ray luminosity function of CTN-AGN in the local universe is described as following $$\frac{d\Phi^{\mathrm{CTN}}(L_X,z=0)}{d \log L_X}  =A\bigg[\bigg(\frac{L_X}{L_*}\bigg)^{\gamma_1} + \bigg(\frac{L_X}{L_*}\bigg)^{\gamma_2}\bigg]^{-1}$$ where A is the normalization of the luminosity function, $L_*$ is the break luminosity, and $\gamma_1$ and $\gamma_2$ are the slopes of the connected power-laws. The luminosity function outside the local universe follows a luminosity and redshift dependent evolution:
$$ \frac{d\Phi^{\mathrm{CTN}}(L_X,z)}{d \log L_X} =\frac{d\Phi^{\mathrm{CTN}}(L_X,z=0)}{d \log L_X}e(L_X,z) $$ where $e(L_X,z)$ is the evolutionary term, which describes the redshift dependence of the luminosity function following equation 16 of \citet{2014ApJ...786..104U}.

\citet{2003ApJ...598..886U} introduced the absorption function $f_{abs}$ which is the probability distribution function defined between $\log N_{\rm H}\ (\rm cm^{-2})=20-26$. The function is normalized to 1 between $\log N_{\rm H} \ (\rm cm^{-2})=20-24$ as follows,
$$ \int_{20}^{24} f_{abs}(L_X, z;  N_{\rm H}) d\log N_{\rm H} = 1 $$ 
As shown in Figure \ref{fig:z_hr_dist}, the amount of absorption in the column density range between $\log N_{\rm H}\ (\rm cm^{-2})=20-22$ cannot be distinguished with the HR used in this analysis. The absorption function was modified by combining the bins between $\log N_{\rm H}\ (\rm cm^{-2})=20-22$  together. The definition is as follows, 

\footnotesize
\begin{equation}
        f_{abs}(L_X, z;  N_{\rm H}) = \left\{\begin{array}{cc} 
            \frac{1-\psi(L_X,z)}{2} & [20\leq\log N_{\rm H} < 22] \\
            \frac{1}{1+\epsilon}\psi(L_X,z) & [22\leq\log N_{\rm H} < 23]\\
            \frac{\epsilon}{1+\epsilon}\psi(L_X,z) & [23\leq\log N_{\rm H} < 24]\\
            \frac{f_{\mathrm{CTK}}}{2}\psi(L_X,z) & [24\leq\log N_{\rm H} < 26]
                                            \end{array} \right. 
\end{equation}
\normalsize

where $\psi(L_X,z)$ is the fraction of CTN-AGN with $\log N_{\rm H}\ (\rm cm^{-2})\geq22$ and $\epsilon$ is the ratio of AGN with $\log N_{\rm H}\ (\rm cm^{-2})=23-24$ to those with $\log N_{\rm H}\ (\rm cm^{-2})=22-23$, and $f_{CTK}$ is the relative number of Compton-thick AGN (CTK-AGN) relative to obscured CTN-AGN, which is fixed to 1. The constraints on $f_{CTK}$ are discussed in Section~\ref{sec:des:ctagn}  

The redshift and luminosity dependence of the absorption function can be described with the function $\psi(L_X,z)$. The obscured fraction in the local universe shows a linearly decreasing dependence on the X-ray luminosity:
\begin{equation}
    \begin{split}
        \psi(L_X,z)=& \mathrm{min}[\psi_{max},\ \mathrm{max}[\psi_{43.75}(z) \\ 
         & -\beta(\log L_X-43.75), \psi_{min} ]]    
    \end{split}
    \label{eq:psi}
\end{equation}
where $\psi_{min}$ and $\psi_{max}$ is the minimum and maximum obscured fraction of CTN-AGN, which is determined to be 0.2 and 0.84, respectively \citep{2014ApJ...786..104U}. The maximum obscured fraction in \citet{2014ApJ...786..104U} was set to 0.84 since larger values of $\psi$ will return negative probability for the absorption function in the $\log N_{\rm H}\ (\rm cm^{-2})=20-21$ bin. Our modification of the absorption function in the $\log N_{\rm H}\ (\rm cm^{-2})=20-22$ removes this effect hence larger values of $\psi_{max}$ are allowed as discussed in the later sections. $\beta$ is the slope of the decrease which is set to 0.24 \citep{2014ApJ...786..104U} and $\psi_{43.75}(z)$ is the obscured fraction at $\log L_X\ (\rm erg\ s^{-1})=43.75$. 

The redshift evolution of the obscured fraction $\psi(L_X,z)$ is described by $\psi_{43.75}(z)$
\begin{equation}
    \psi_{43.75}(z) = \left\{\begin{array}{cc} 
                            \psi_{43.75}^0(1+z)^{a1}& \hspace{5mm} z < 2.0 \\
                            \psi_{43.75}^0(1+2)^{a1}=\psi_{43.75}^2 & \hspace{5mm} z \geq 2.0 \\
                          \end{array} \right.  
    \label{eq:psi4375}
\end{equation}
where $\psi_{43.75}^0$ is the obscured fraction at $\log L_X\ (\rm erg\  s^{-1})=43.75$ in the local universe. It was determined to be $0.43\pm 0.03$ (\citet{2014ApJ...786..104U}). The redshift dependence parameter $a_1$ is determined to be $0.48\pm 0.05$. We assumed a constant $\psi_{43.75}(z=2) = \psi_{43.75}^2$ in the redshift range above redshift 2.

It should be pointed out that $\psi_{43.75}(z)$ is a parameter used to control the evolution of the obscured fraction $\psi(L_X,z)$ and is not limited to between $\psi_{min}$ and $\psi_{max}$.  As defined in \ref{eq:psi}, $\psi_{43.75}(z)$ is equal to $\psi(43.75,z)$ only when $\psi_{43.75}(z)\leq\psi_{max}$, beyond that would suggest that $\psi(43.75,z)$ has saturated at $\psi_{max}$ with no further evolution with redshift but the obscured fraction of higher luminosity AGN can still be lower. Therefore, the obscured fraction was estimated with $\psi(L_X,z)$ (see Section \ref{sec:res:intobs}).  

\subsection{Survey Area Function} \label{sec:analy:surveyvol}

In order to estimate the obscured fraction at high redshifts, it is important to evaluate the dependence of the effective survey volume as a function of luminosity, redshift, and amount of obscuration. The survey area that is sensitive enough to detect obscured AGN can be smaller than that for unobscured AGN at the same intrinsic luminosity.  The survey area function  $\Omega(L_X,z,\log N_{\rm H})$ was estimated as a function of AGN 2-10 keV luminosity, redshift, and $\log N_{\rm H}$ using XSPEC \citep{1996ASPC..101...17A}.  The same phenomenological AGN model, RMF and ARF calibration files of the PN-detector used to estimate the $N_{\rm H}$, and the survey area curve defined in Section \ref{sec:analy:area} are considered.

\begin{figure}[ht!]
    \centering
    \epsscale{1.1}
    \plotone{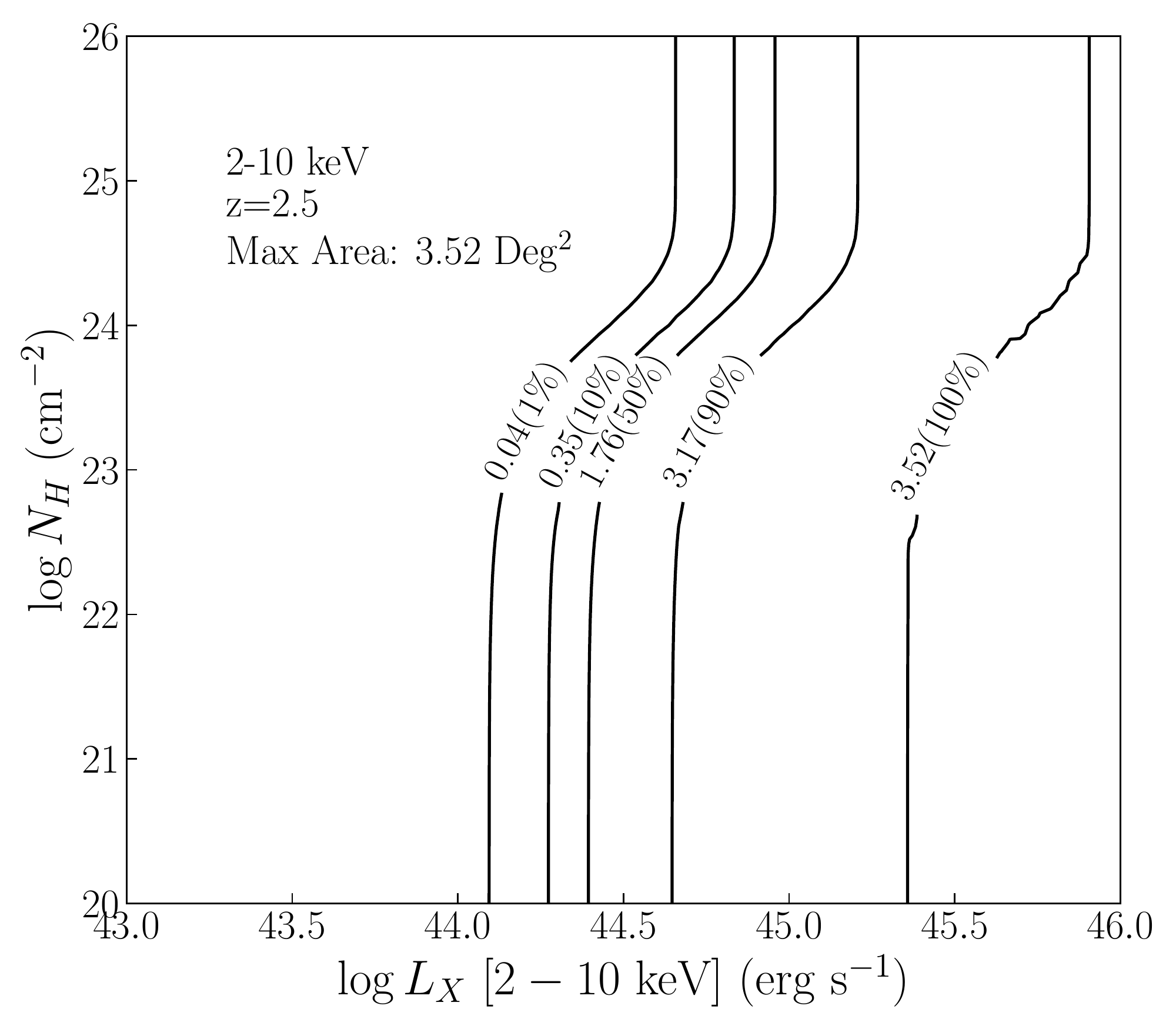}
    \caption{Survey area of the 2-10 keV band sample at redshift 2.5 as a function of $\log L_X$ and $\log N_{\rm H}$. The contours represent survey area at 1\%, 10\%, 50\%, 90\%, and 100\% of the total survey area (3.52 $\rm Deg^2$).}
    \label{fig:surveyarea}
\end{figure}

Figure \ref{fig:surveyarea} shows the survey area at redshift 2.5 as a function of $\log L_X$ and $\log N_{\rm H}$ in the  2-10 keV band. At a fixed X-ray luminosity, the accessible survey area decreases with increasing column density due to increasing X-ray extinction.  Above $\log N_{\rm H}\ (\rm cm^{-2})=24$, the area becomes constant because the scattered and reflected components dominate the spectra.

\subsection{Maximum Likelihood Fitting} \label{sec:analy:mlf}

Maximum-likelihood fitting was performed to estimate the obscured fraction of quasars using the high redshift AGN sample. The sample for the maximum-likelihood fitting was constructed from the 304 $z=2-5$ AGN detected in the 2-10 keV band. For each source, the survey area based on $\log L_X$, z, and $\log N_{\rm H}$ was calculated in order to evaluate the possibility to detect that AGN. One AGN was removed since the corresponding survey area is zero. For the AGN detected only in the 2-10 keV band, we assign the column density of $\log N_{\rm H}\ (\rm cm^{-2}) = 23.5$.

The maximum likelihood (ML) method is a parametric fitting method, which uses the observed parameters of each object without binning.  The likelihood function ($\mathcal{L}$) is generally defined as the product of all probability densities ($P$) in the sample. The probability density is defined as the probability of finding the \textit{i}-th object with $\log N_{\rm H}^i$ at $\log L_X^i$ and $z^i$ as $P_i$ $$    P_{i}=\frac{f_{abs}(L_X^i, z^i;\log N_{\rm H}^i)\Omega(L_X^i, z^i,\log N_{\rm H}^i)}{\int_{20}^{24} f_{abs}(L_X^i, z^i; \log  N_{\rm H})\Omega(L_X^i, z^i, \log N_{\rm H}) d\log N_{\rm H}}$$  
where $f_{abs}$ and  $\Omega$ is the absorption function and survey area function, respectively. 

The likelihood function in the logarithmic form $\mathcal{M}$ is then defined as the sum of the logarithmic probability density of the \textit{i}-th AGN within the fitting sample $$\mathcal{M}(\psi_{43.75}^2,\epsilon)=-2 \sum_{i} \ln{P_{i}(\psi_{43.75}^2,\epsilon)}.$$ The best-fit parameters are obtained by minimizing the likelihood function over the parameter space of interest. In our case, we set $\epsilon$ and $\psi^2_{43.75}$ as free parameters. The 1-sigma uncertainty of the best-fit parameters was estimated based on the range where the log likelihood value changes from minimum by one.

\section{Results} 

\subsection{The Obscured Fraction in the Luminous End} \label{sec:res:intobs}

The maximum likelihood fitting was performed in two ways, 1) by fitting $\psi_{43.75}^2$ with fixed $\epsilon=1.7$, which is the parameter determined in the local universe \citep{2014ApJ...786..104U} (hereafter we refer to as ``1D") and 2) by fitting both of $\psi_{43.75}^2$ and $\epsilon$ simultaneously (hereafter ``2D"). 

At first, the fitting was performed with the maximum obscured fraction ($\psi_{max}$) set to be 0.84 following \citet{2014ApJ...786..104U}. The results suggested that the fitting is affected by the choice of $\psi_{max}$. Thus, the maximum obscured fraction $\psi_{max}$ was set to 0.99 instead. The best-fit results from the 1D case and 2D cases are shown in Table \ref{tab:fitres}. 

\begin{figure}[!htb]
    \centering
    \plotone{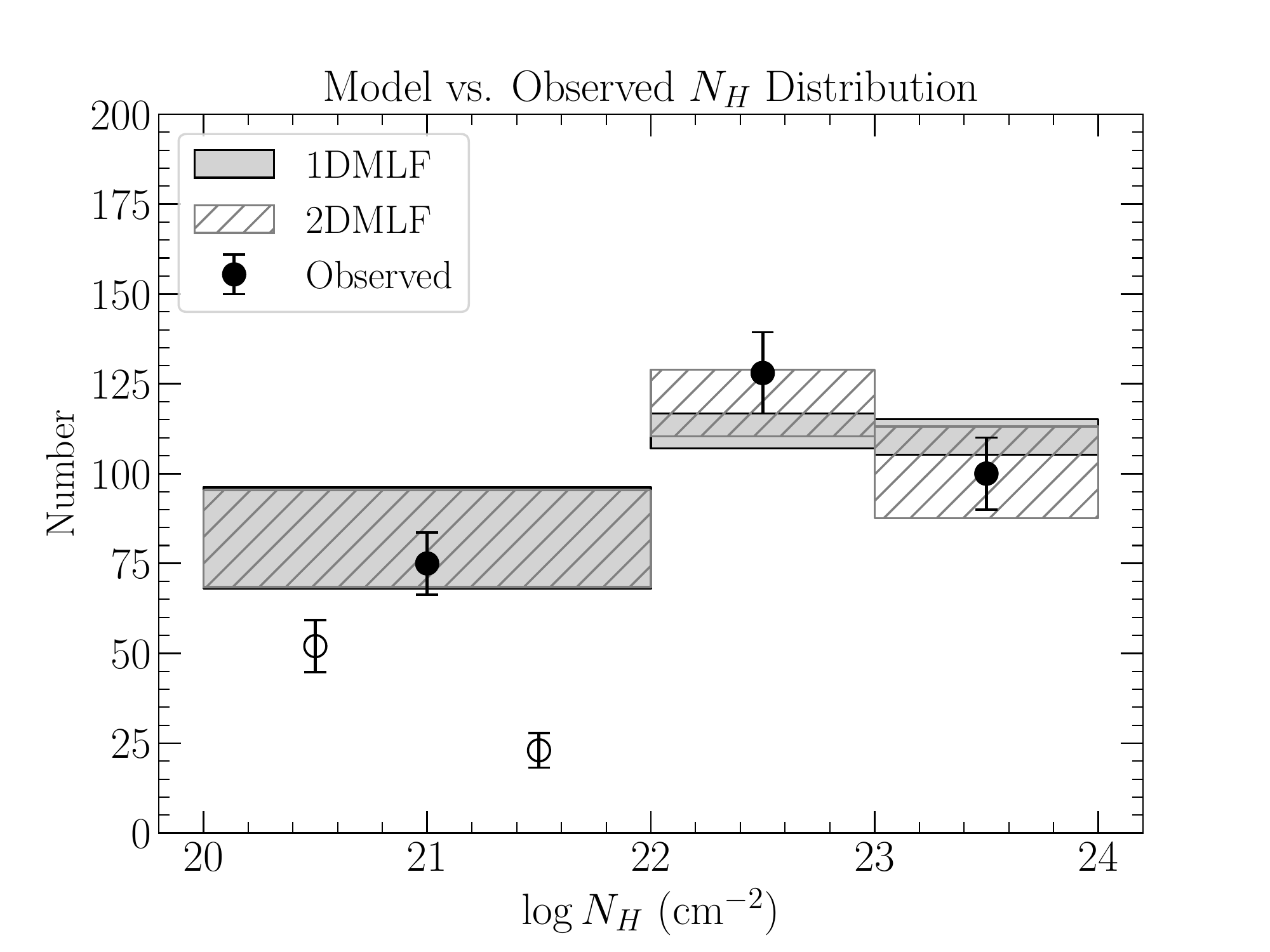}
    \caption{Expected $\log N_{\rm H}$ distribution determined from the best fit models of 1D and 2D ML-fit shown filled gray and hatched gray boxes, respectively. The upper and lower edge of each box represents the upper-lower limit of the models. The observed $N_{\rm H}$ distribution of the 2-10 keV band detected high redshift AGN is shown with black round symbols.  AGN with $\log N_{\rm H}\ (\rm cm^{-2})<22$ cannot be distinguished using the HR and are combined into a single bin. Open round symbols show the unbinned observed number of AGN with $\log N_{\rm H}(\rm cm^{-2})<22$. AGN detected only in the 2-10 keV band are added to the $\log N_{\rm H}\ (\rm cm^{-2})=23-24$ bin.  The expected distributions from 1D and 2D fits are scaled by a factor of 1.72 and 1.68 so that the total number of CTN-AGN is consistent with observation. }
    \label{fig:bf_nhdist}
\end{figure}

\begin{figure*}[!htb]
    \centering
    \epsscale{1}
    \plottwo{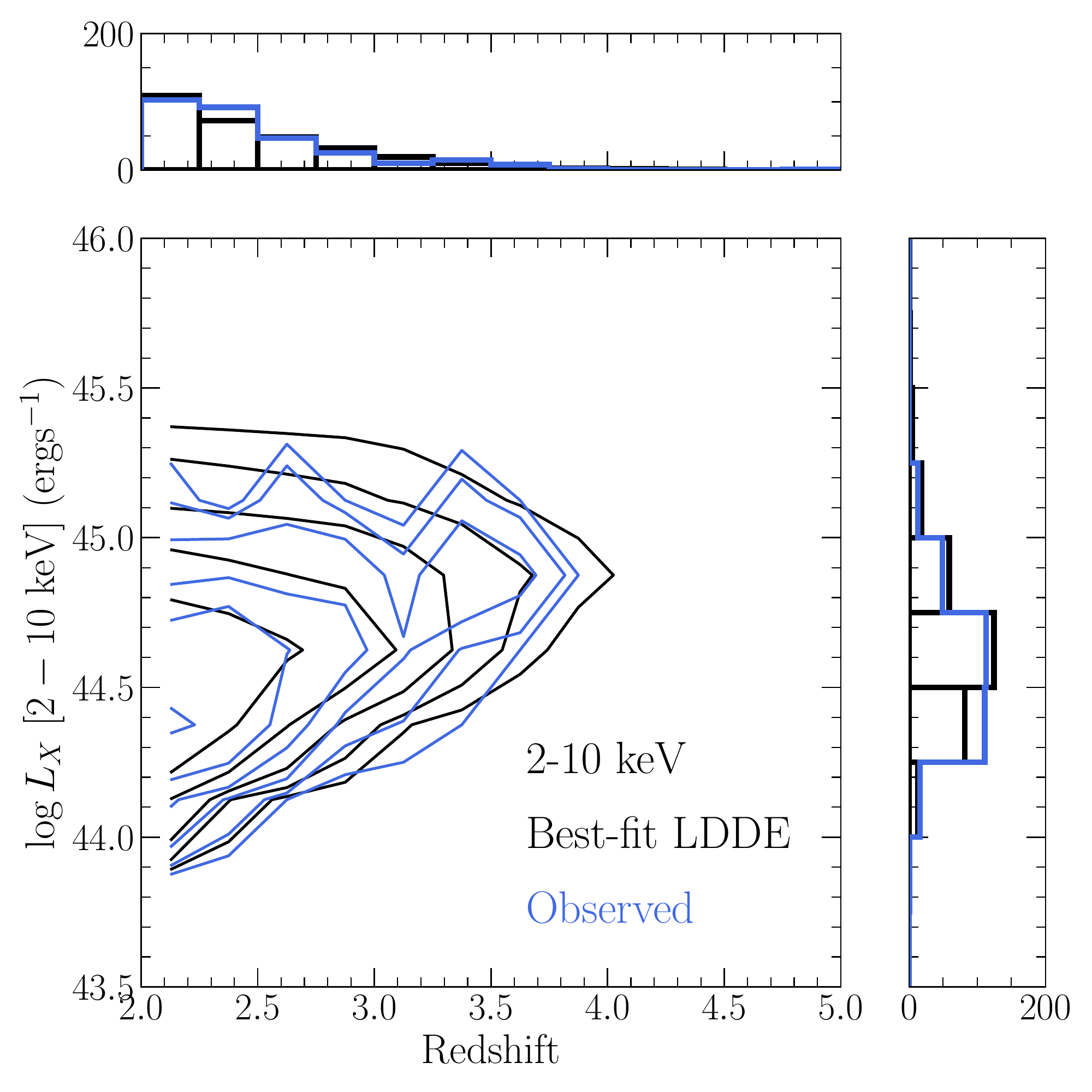}{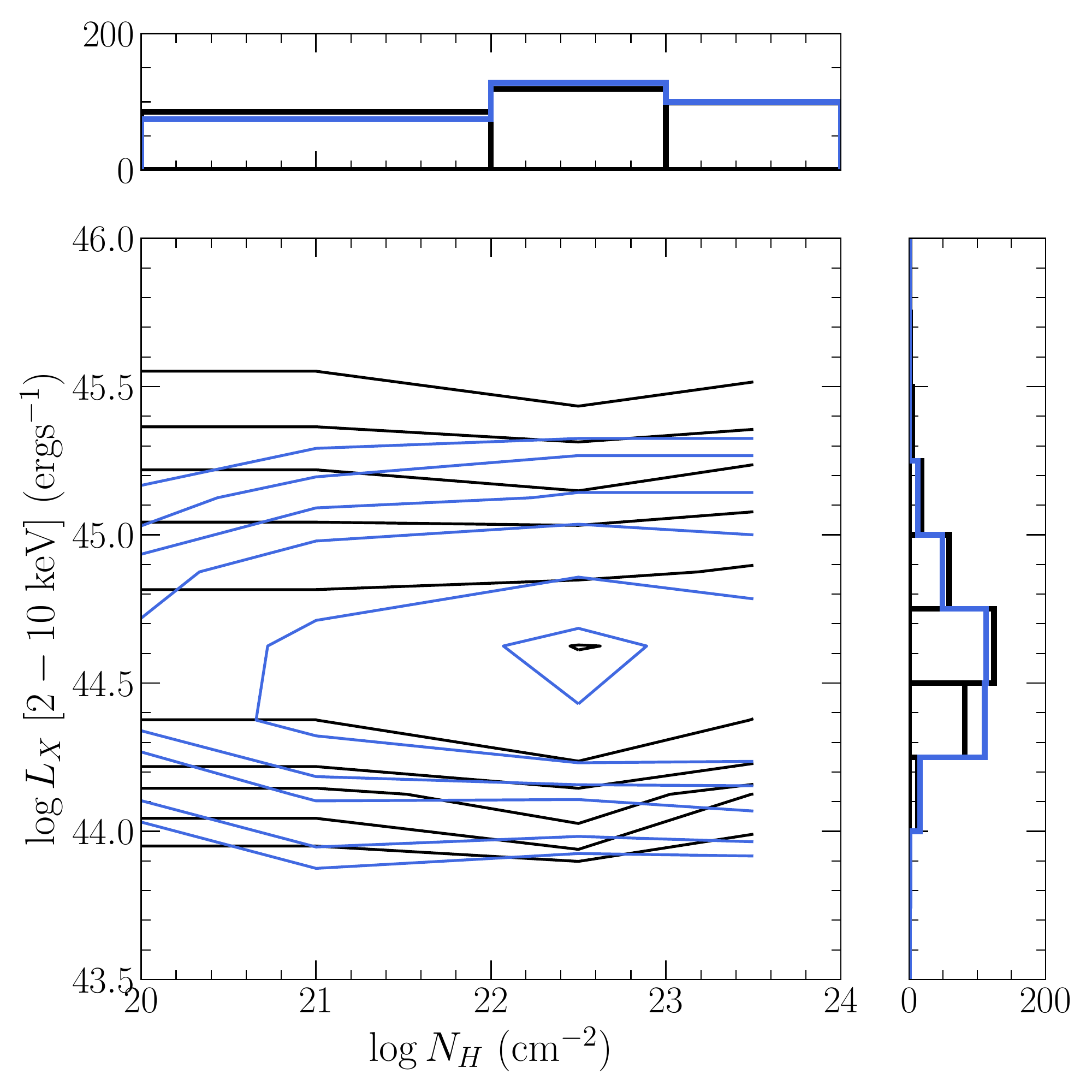}
    \caption{(Left) Redshift and 2-10 keV luminosity distribution. The 2-10 keV  detected high redshift AGN sample distribution is shown in blue while the expected distribution from the 2D ML fit is shown in black.  The expected distribution was scaled by a factor of 1.68 so that the total number of CTN-AGN is consistent with observation. (Right) Same as left but showing the $\log N_{\rm H}$ and 2-10 keV luminosity distribution. In both panels, the top and right histograms show the projected distribution of each axes.}
    \label{fig:bf_lxzdist}
\end{figure*}

Figure \ref{fig:bf_nhdist} shows the observed $\log N_{\rm H}$ distribution of the 2-10 keV band detected AGN. The comparison suggests that the fitting results from the 2D case reproduce the observed distribution better than 1D. Figure \ref{fig:bf_lxzdist} shows the predicted distribution of redshift, 2-10 keV luminosity, and $\log N_{\rm H}$  based on the best-fitted parameters from the 2D ML-fit. The predicted distributions reproduce the observed distribution well. Therefore, the results of the 2D-ML fit are adopted for further discussion.

It should be pointed out that the parameter $\psi_{43.75}$ represents the obscured fraction of moderately luminous quasars with $\log L_X\ (\rm erg\ s^{-1})=43.75$, however, the current sample covers luminosity range of $\log L_X\ (\rm erg\ s^{-1})\sim44-45$. The obscured fraction of luminous quasars was estimated by calculating the intrinsic number of luminous obscured CTN-AGN. The expected number of AGN in each $\log N_{\rm H}$ bin is calculated with the following equation:
\begin{equation}
    \begin{split}
    N= & \int\int\int f_{abs}(\log L_X, z, N_{\rm H})\frac{d\Phi(\log L_X, z)}{d \log L_X} \\
        &\Omega(1+z)^3 d_A(z)^2 \frac{d\tau}{dz} d \log L_X\ dz\ d\log N_{\rm H}
    \end{split}
    \label{eq:nagn}
\end{equation}
where $f_{abs}$, $d\Phi$, $\Omega$, $d\log L_X$, $d_A$, and $\tau$ are the absorption function, 2-10 keV luminosity function, survey area, angular size distance, and look-back time, respectively. The integration limits are between the redshift, luminosity, and column density of interest. For the expected number of AGN, the survey area $\Omega$ is replaced with the survey area function $\Omega(L_X,z,\log N_{\rm H})$.

\begin{deluxetable}{CCC}[!htb] \label{tab:fitres}
\tablecaption{Maximum Likelihood Fitting Results}
\tablewidth{0pt}
\tablehead{
    \colhead{Parameter} & \colhead{1D} & \colhead{2D} 
}
\startdata
\epsilon & 1.7\ (\rm fixed) & 1.4\pm 0.3 \\
\hline
\psi_{43.75}^2 & 1.01_{-0.04}^{+0.03} & 0.99_{-0.03}^{+0.04} \\
\enddata
\end{deluxetable}

Based on the best-fit parameters from the 2D ML fit, the obscured fraction of the luminous quasars with $\log L_X\ (\rm erg\  s^{-1})>44.5$ is estimated to be $0.76_{-0.03}^{+0.04}$. The lower limit of the luminosity integration was set to $\log L_X\ (\rm erg\ s^{-1})=44.5$, as it corresponds approximately to the break in the hard X-ray luminosity function at redshift 2. 

Our best-fit obscured fraction is the largest compared to the obscured fraction which used the same model and in the same luminosity range. \citet{2012ApJ...758...49H} estimated the obscured fraction of quasars with $\log L_X\ (\rm erg\ s^{-1})=44-45$ at redshift 3-5 to be $0.54^{+0.17}_{-0.19}$, while the best-fit parameters of \citet{2014ApJ...786..104U} suggest an obscured fraction of $0.50\pm0.09$ above redshift 2. 
 
For comparison with other studies, the obscured fraction of AGN with $\log L_X\ (\rm erg\ s^{-1})=44-45$ based on the 2D ML best-fit parameters is adopted. For CTN-AGN, the obscured fraction is $0.85_{-0.03}^{+0.04}$. If we calculate the obscured fraction with $\log N_{\rm H} (\rm cm^{-2})>23$, the obscured fraction is $0.50_{-0.06}^{+0.07}$.

\subsection{The Slope of the Obscured Fraction on the Luminous End} \label{sec:res:slope}
The maximum likelihood method applied in Section \ref{sec:res:intobs} assumes that the slope of the obscured fraction dependence on luminosity does not depend on redshift. Therefore, the rate at that the obscured fraction increases is the same at all luminosity as long as it has not saturated. Here, the slope of the luminosity dependence on the luminous end at high redshift was investigated by applying the maximum-likelihood method to determine the slope parameter $\beta$ assuming that $\psi_{43.75}^2=0.73$ following \citet{2014ApJ...786..104U} and $\epsilon=1.4$ following the 2D ML-fit results. 

The best-fit slope determined using the ML fit is $-0.09_{-0.03}^{+0.04}$, which suggests an obscured fraction of luminous CTN-AGN with $\log L_X\ (\rm erg\ s^{-1})=44-45$ is $0.78\pm0.02$. The best-fit slope suggests that the obscured fraction shows almost no luminosity dependence. However, it should be noted that the luminosity coverage is limited to between  $\log L_X\ (\rm erg\ s^{-1})\sim44-45$, and the obscured fraction at higher luminosity can not be constrained.

\subsection{Systematic Uncertainties in the Analysis} \label{sec:des:sys}

First, systematic uncertainties arise from the photon index and the reflection strength assumed in the phenomenological AGN model which is used to calculate the column density using the hardness ratio. 

If we assume a photon index of 1.7 with the same reflection strength, the obscured fraction for luminous quasars with $\log L_X\ (\rm erg\ s^{-1})=44-45$ is estimated to be $0.77_{-0.04}^{+0.03}$, which is approximately $9\%$ lower than when assuming a photon index of 1.8 ($0.85_{-0.03}^{+0.04}$). Assuming a photon index of 1.8 with a stronger reflection strength of 1.3 reduces the obscured fraction to $0.82\pm0.03$ for luminous quasars, corresponding to a systematic change of approximately 3\%.  The assumption of the photon index and the reflection strength does not strongly affect the estimate of the column density ratio ($\epsilon$) since the results are consistent with each other within uncertainties.

Second, the obscured fraction based on a hard X-ray selected AGN sample could suffer from a bias in which the 2-10 keV count-rates close to the detection limit are larger than the true count-rates due to statistical fluctuations of the photon count rates (Eddington bias). As a result, hard X-ray selected AGN samples may have harder hardness ratios and as a result, larger column densities overall. This statistical fluctuation may also affect intrinsically unobscured AGN which makes them show large column density consistent with obscured AGN due to positive fluctuation in the 2-10 keV band.

Last, the estimates of the column density and luminosity may also be affected by the uncertainty in the photometric redshift. Contamination from low-redshift AGNs ($z<2$) can also affect the best-fit results. We estimate the contamination rate from low-redshift AGN ($z_{\rm spec}<2$) based on the fraction of outlier spectroscopically confirmed quasars above redshift two (AGN above the red-dashed line with $z_{\rm phz}>2$ as shown in Figure \ref{fig:photozcompare}) over all AGN above redshift two ($z_{\rm phz}>2$) to be 31\%.

\section{Discussion}  \label{sec:des}
\subsection{Redshift Dependence of the Obscured Fraction} \label{sec:des:obsf}

\begin{figure}[!htb]
    \centering
    \epsscale{1.1}
    \plotone{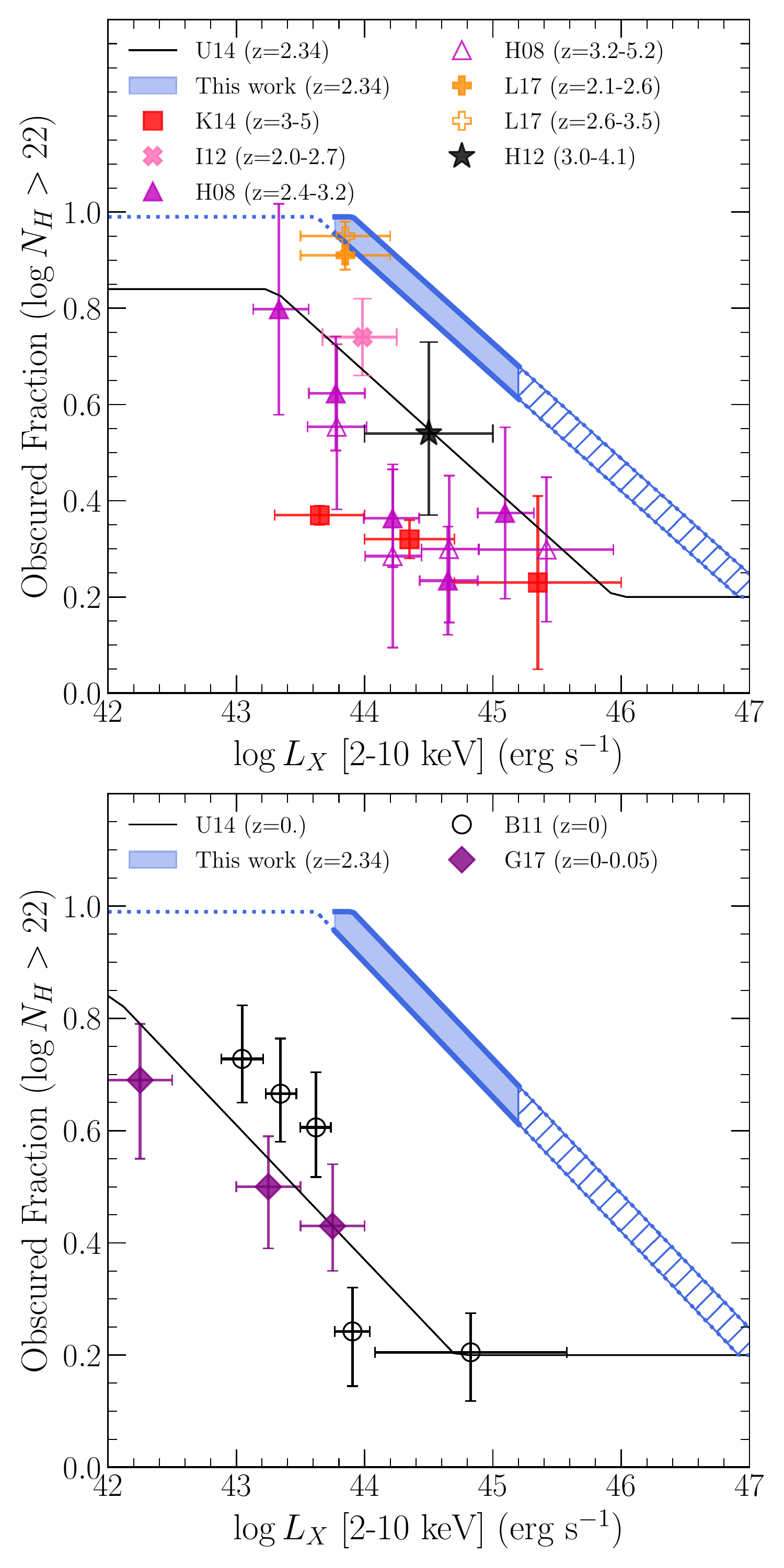}
    \caption{Obscured fraction as a function of 2-10 keV luminosity based on the 2D ML best-fit parameters blue shaded area, the lower and upper limit is shown with thick blue line. The luminosity range not covered by our sample are shown with the hatched area. The top panel shows the results at $z>2$ compared with \citet{2008A&A...490..905H} (H08; opened and closed magenta triangles), \citet{2012A&A...546A..84I} (I12; pink cross), \citet{2012ApJ...758...49H} (H12; black star), \citet{2014MNRAS.445.1430K} (K14; red squares), and \citet{2017ApJS..232....8L}(L17; opened and closed plus symbols), and \citet{2014ApJ...786..104U} (U14, thin black line), respectively. The bottom panel shows the results at low redshift compared with \citet{2011ApJ...728...58B} (B11; open black circles), \citet{2017MNRAS.469.3232G} (G17;purple diamonds), and \citet{2014ApJ...786..104U}(U14; solid black line), respectively. }
    \label{fig:compare_lxobsfrac}
\end{figure}

Using the best-fit parameters from the 2D-ML method, we calculate the obscured fraction of CTN-AGN as a function of 2-10 keV luminosity by assuming the luminosity dependent obscured fraction presented in Section \ref{sec:analy:lddemod}. Figure \ref{fig:compare_lxobsfrac} shows the best-fit obscured fraction based on the 2D ML best-fit parameters compared with previous measurements in the local universe (bottom) \citep{2011ApJ...728...58B, 2014ApJ...786..104U,2017MNRAS.469.3232G} and at $z=2-5$ (top) \citep{2008A&A...490..905H,2012A&A...546A..84I,2012ApJ...758...49H,2014MNRAS.445.1430K,2014ApJ...786..104U,2017ApJS..232....8L}.  Our estimate of the obscured fraction is larger than the obscured fraction in the local universe, which supports the increasing trend in the obscured fraction towards high redshift. At high redshift, our estimate of the obscured fraction at $\log L_X\ (\rm erg\ s^{-1})={43.75}$ is the largest compared with studies in the same redshift range, except for \citet{2017ApJS..232....8L}, in which the obscured fraction in the same redshift range and with $\log L_X\ (\rm erg\ s^{-1})=43.5-44.2$ is determined to be $0.91\pm0.03$. The luminosity coverage of their sample is below the luminosity range of our sample but consistent with our estimate of the obscured fraction at $\log L_X\ (\rm erg\ s^{-1})=43.75$ if we extrapolate the fraction toward lower luminosity without the maximum limit of 0.84 and compare with the obscured fraction of AGN with $\log N_{\rm H}\ (\rm cm^{-2})>22$. 

\begin{figure}[!htb]
    \centering
    \plotone{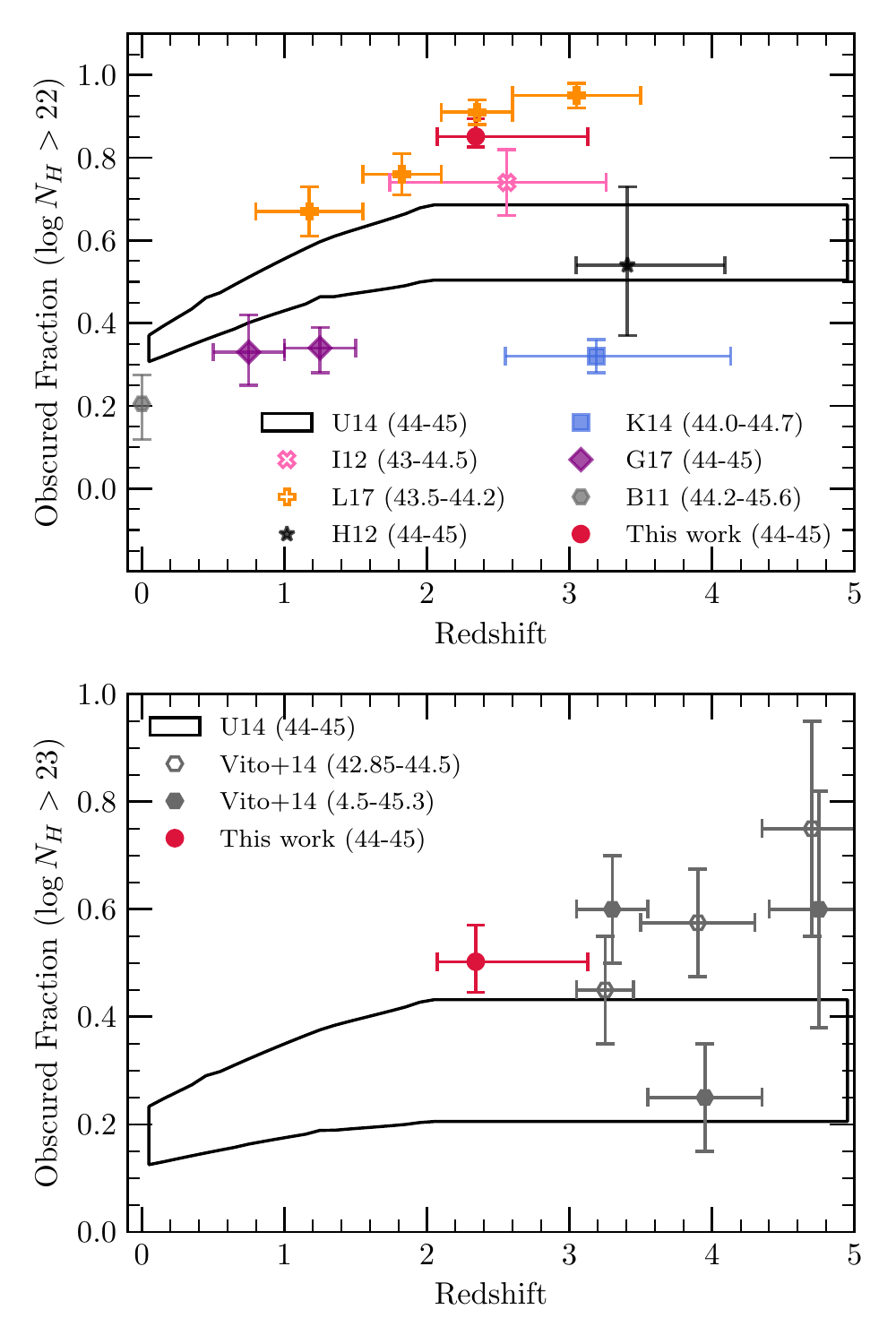}
    \caption{The obscured fraction at different redshifts compared with the obscured fraction of AGN with $\log L_X\ (\rm erg\ s^{-1})=44-45$ from the 2D ML best-fit parameters shown with the red round symbol. (Top) Obscured fraction of AGN with $\log N_{\rm H}\ (\rm cm^{-2})>22$ with redshift.  The result is compared with \citet{2011ApJ...728...58B}(B11; grey hexagon), \citet{2012A&A...546A..84I}(I12; pink cross), \citet{2012ApJ...758...49H}(H12; black star) \citet{2014MNRAS.445.1430K}(K12; blue square),  \citet{2017ApJS..232....8L}(L17; orange plus), and \citet{2017MNRAS.469.3232G} (G17; purple diamonds), respectively.  (Bottom) The obscured fraction of AGN with $\log N_{\rm H}\ (\rm cm^{-2})>23$ among CTN-AGN. The results are compared with \citet{2014MNRAS.445.3557V} shown with gray hexagons. In both panels, The black solid line shows lower and upper limits of \citet{2014ApJ...786..104U}(U14). The $\log L_X\ (erg s^{-1})$ range are shown in parenthesis. }
    \label{fig:compare_zobsfrac}
\end{figure}

Upper and lower panels of Figure \ref{fig:compare_zobsfrac} show the obscured fraction of AGN  with $\log N_{\rm H}\ (\rm cm^{-2})>22$ and $>23$, respectively. The obscured fraction is larger than the obscured fraction in the same luminosity range from \citet{2014ApJ...786..104U} (black solid line). The larger obscured fraction suggests that the evolution of the obscured fraction needs to be stronger at $z<2$.  In order to reconcile with our estimate of the obscured fraction, the evolution parameter (a1) of the obscured fraction must be $0.75_{-0.08}^{+0.10}$. However, this will systematically increase the obscured fraction at $z<2$. Alternatively, the obscured fraction may still evolve above redshift 2 and the obscured fraction of high luminosity AGN saturates at a higher redshift than that of low luminosity AGN. 

\citet{2014MNRAS.445.3557V} estimated the obscured fraction of $\log N_{\rm H}\ (\rm cm^{-2})>23$ and $\log L_X\ ( \rm erg\ s^{-1}) \geq 43$ AGN at redshift 3-5 to be $0.54\pm 0.05$. Our sample resides at a lower redshift than that of \citet{2014MNRAS.445.3557V}. However, our model assumes no evolution of the obscured fraction above redshift two. The obscured fraction of AGN based on the same definition and redshift and luminosity range was estimated to be $0.57\pm0.05$ based on the best-fit parameters of the 2D ML fit. This is consistent with that of \citet{2014MNRAS.445.3557V} within 1$\sigma$ uncertainty.
 
\citet{2017ApJS..232....8L} also examined the obscured fraction using AGN samples from the C-COSMOS legacy \citep{2011ApJ...741...91C,2016ApJ...819...62C} combined with the Chandra deep field south 7M catalog (CDF-S, \citealt{2017ApJS..228....2L}). The obscured fraction of AGN with $\log N_{\rm H} (\rm cm^{-2})=22-23$ among those with $\log N_{\rm H} (\rm cm^{-2})<23$. For AGN with $\log L_X\ (\rm erg\ s^{-1})=44.1-44.9$ at redshift 2-3, the fraction is $0.67\pm0.07$. The obscured fraction estimated based on the same definition between $\log L_X\ (\rm erg\ s^{-1})=44-45$ using the 2D ML best-fit parameters  is $0.70_{-0.01}^{+0.05}$ which is consistent to the above value within 1$\sigma$ uncertainty.

Recently, \citet{2022arXiv220603508G} estimated the obscured fraction of AGNs as a function of redshift by constructing an ISM model based on ALMA data. Our results are consistent with those of \citet{2022arXiv220603508G} for AGN with $\log N_{\rm H}\ (\rm cm^{-2})>23$ with $\log L_X\ (\rm erg\ s^{-1})\sim44.$ but higher than the $1\sigma$ uncertainties for the obscured fraction of AGN with $\log N_{\rm H}\ (\rm cm^{-2})>22$ at the same luminosity. The larger obscured fraction in our study in the case of $\log N_{\rm H}\ (\rm cm^{-2})>22$ may be due to the difficulty in separating unobscured AGN from mildly obscured AGN through the hardness ratios.

\subsection{Implications of an Increasing Obscured Fraction}  \label{sec:des:imp}

The increasing trend in the obscured fraction has strong implications on the physical structure and evolution in the nuclear environment of the AGN. The larger obscured fraction at high redshift implies that there is a larger amount of obscuring material within the nuclear, circumnuclear, or host galaxy scale compared to AGN in the local universe.

The trend may be a direct result of the evolution in the host galaxy scale properties of interstellar matter (ISM). Observations of massive galaxies at high redshift show that they are more compact \citep{2014ApJ...788...28V} and have higher gas fractions \citep{2010Natur.463..781T,2013ARA&A..51..105C} compared to those in the local universe with the same stellar mass. This may suggest that the gas density is higher on all spatial scales of the host galaxy compared to those in the local universe.  As a result, the higher occurrence of obscuration in the host galaxy or circumnuclear region may explain the increasing trend of the obscured fraction of CTN-AGN \citep{2017MNRAS.465.4348B,2017MNRAS.464.4545B,2019A&A...623A.172C,2008MNRAS.385L..43F,2022arXiv220603508G} and may explain the Compton-thick fraction \citep{2020A&A...636A..37D}. Due to the denser ISM and metal abundance of gas in the circumnuclear region and host galaxy at high redshift, AGN feedback can be less efficient in clearing sight-lines since gas can easily cool and replenish the obscuring material \citep{2019MNRAS.487..819T}. The decrease in the obscured fraction from high to low redshift may then be explained due to gas consumption by star-formation and AGN accretion \citep{2014MNRAS.442.2304H}. Feedback by AGN-driven winds further reduces the obscuration within and possibly beyond the nuclear region, as spectroscopic observations of high redshift AGN show that AGN can drive winds with velocities from several hundred up to a thousand kilometers per second, such outflows can reach out to several kiloparsecs beyond the nuclear region \citep{2016A&A...586A.152C,2017A&A...599A.123N,2020ApJ...894...28D}.  

Another possibility is that the trend in the obscured fraction is driven by the triggering of AGN by major mergers. Galaxy merger simulations suggest that luminous quasars may go through an evolutionary sequence. In this scenario, strong gravitational effects by a merging event funnel large amounts of gas and dust towards the nuclear region, which results in a heavily obscured nuclear activity \citep{2006ApJS..163....1H,2008ApJS..175..390H,2009ApJ...696..891H}. During the obscured quasar phase, obscuration with a large column density ($\log N_{\rm H}\ (\rm cm^{-2}) > 22$) as well as near Eddington limited accretion are induced.  This phase is expected to last as long as 10 times the following blow-out phase, in which strong winds driven by the AGN blows out the obscuring material leaving an unobscured quasar at the end of the sequence. It is possible that the fraction of AGN triggered by major merger is higher at higher redshifts; observations of merging galaxies in the local universe show that they are heavily obscured compared to those triggered in other processes \citep{2017MNRAS.468.1273R,2021MNRAS.506.5935R}.

Lastly, the trend in obscured fraction with redshift may be compatible in the context of radiation pressure on nuclear dust where after exceeding an $N_{\rm H}$ critical effective Eddington ratio, the dust in the nuclear region is blown out \citep{2006MNRAS.373L..16F,2008MNRAS.385L..43F,2009MNRAS.394L..89F,2015MNRAS.451...93I}. In contrast to the merger-driven scenario, mergers are not a prerequisite in order to drive strong outflows. In this model, nuclear obscuration larger than $N_{\rm H}>5\times10^{21} \rm cm^{-2}$ occurs from long-lived dust clouds near the AGN while milder obscuration is due to dust lanes outside the SMBH gravitational sphere of influence and independent of the AGN Eddington ratio. At high redshift, accretion activity occurs with higher Eddington ratios than AGN in the local universe \citep{2012ApJ...761..143N,2015MNRAS.447.2085S}. The higher average Eddington ratio suggests more AGN are in a blow out phase and less affected by nuclear obscuration. One possibility to explain the increasing trend of the obscured fraction is that the dust abundance in the nuclear region for high redshift AGN is lower than those in the local universe \citep{2009MNRAS.394L..89F}. 

\subsection{High-redshift Compton-Thick AGN} \label{sec:des:ctagn}

Compton-thick AGN (CTK-AGN) are heavily obscured AGN with $\log N_{\rm H}\ (\rm cm^{-2})>24$. Due to the heavy obscuration, the AGN X-ray continuum emission is strongly suppressed, thus their detection requires the selection of hard X-rays above restframe 10 keV. 

Among the high-redshift AGN sample, 53 AGN were detected only in the 2-10 keV band and have lower limits for $\log N_{\rm H}$ larger than $\log N_{\rm H}\ (\rm cm^{-2})\geq23$. Considering the band shifting effect, non-detection in the 0.5-2 keV band suggests their AGN X-ray continuum below restframe 6-8 keV is heavily suppressed. Therefore, the 53 AGN may be considered as CTK-AGN candidates at high redshift.

Assuming the phenomenological AGN model, the survey area function presented in section \ref{sec:analy:surveyvol}, and using the absorption function in section \ref{sec:analy:lddemod} with $f_{\rm CTK}=1$, the expected number of CTK-AGN detected with $\log N_H\ (cm^{-2})=24-26$ is $25.61_{-1.61}^{+0.95}$. This may suggest that roughly half of the 2-10 single band detected AGN may be CTK-AGN candidates.   However, the column density determination based on the 0.5-2.0 and 2.0-10.0 keV bands HR cannot discriminate heavily obscured CTN-AGN and CTK-AGN in the sample as shown in Figure \ref{fig:z_hr_dist}. In addition, the adoption of physical torus models, X-ray spectral analysis, or secondary tracers of AGN luminosity are generally required to reliably confirm the CTK nature \citep{2017ApJS..233...17R}. A full discrimination of the CTK population is beyond the scope of this paper.

\begin{figure*}[!htb]
    \centering
    \plotone{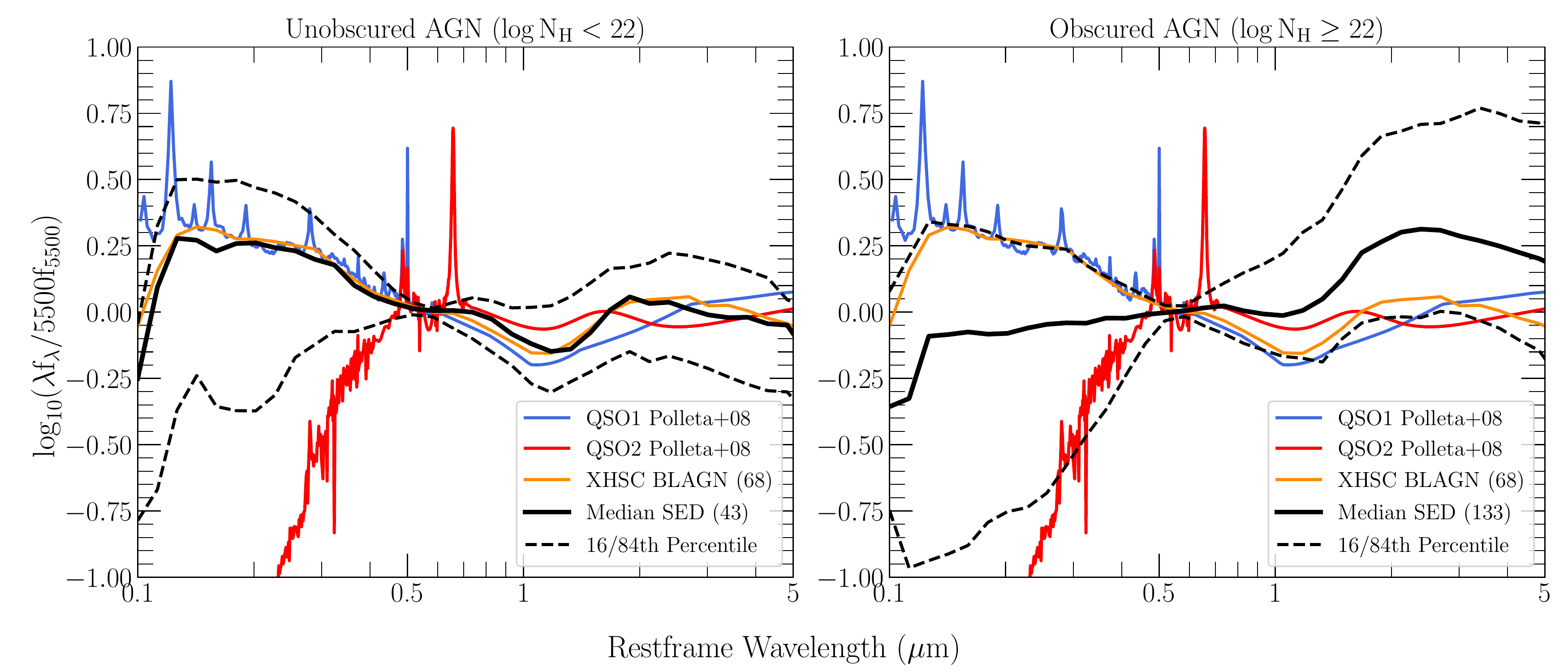}
    \caption{(Left) Median SED of X-ray unobscured AGN with $\log N_{\rm H} \ (\rm cm^{-2})< 22$ (black solid line). The 16th and 84th percentile SED are plotted with black dashed lines. (Right)  Same as left but for X-ray obscured AGN with $\log N_{\rm H} \ (\rm cm^{-2})\geq 22$ (black solid line). In both diagrams, the median SED of high redshift BL-AGN is shown in orange while type-1 QSO and type-2 QSO SED from \citealt{2007ApJ...663...81P} are shown in blue and red, respectively. The number in parenthesis represents the number of objects used to produce the SED. All SEDs are normalized at $\rm5500\AA$.  }
    \label{fig:nhbin_sed}
\end{figure*}

\subsection{The Rest-frame Spectral Energy Distribution} \label{sec:des:sed}

We examine the correspondence between the X-ray obscuration and the SED shapes between UV and IR bands by examining the restframe SED of the of high redshift quasars ($\log L_X\ (\rm erg\ s^{-1})>44.5$) detected in the 2-10 keV band. For each quasar, the SED were constructed by interpolating between the 12-band photometry shifted to restframe and normalized at 5500$\rm \AA$. We also include additional IRAC3 ($\rm 5.8 \mu \mathrm{m}$), IRAC4 ($\rm 8 \mu \mathrm{m}$), and MIPS 24$\rm \mu \mathrm{m}$ bands photometry from the SWIRE dataset \citep{2003PASP..115..897L}. The AGN sample was separated into X-ray unobscured ($\log N_{\rm H}\ (\rm cm^{-2}) < 22$) and X-ray obscured ($\log N_{\rm H} \ (\rm cm^{-2})\geq 22$ ) AGN based on the column density. The median SED of X-ray unobscured and obscured quasars was constructed by median-combining the individual SED of X-ray unobscured and obscured quasars, respectively.

Figure \ref{fig:nhbin_sed} shows the median SED of the high redshift quasars compared with type-1 and type-2 QSO SEDs from \citet{2007ApJ...663...81P} as well as the median SED of high redshift BL-AGN. The median SED of X-ray unobscured AGN is flat similar to the median SED of the BL-AGN. This is consistent with the expected power-law SED of unobscured AGN. On the other hand, the median SED of X-ray obscured AGN shows a redder UV, optical, and near-infrared continuum compared to the BL-AGN. 

Both the X-ray unobscured and obscured AGN show a variety of SED shapes as shown in the 16th and 84th percentile SED distribution. More than 16\% of the obscured AGN have a blue UV continuum similar to the median SED of the BL-AGN, while 16\% of the unobscured AGN are redder than the median SED of the obscured AGN. This suggests that the correspondence between the UV- optical SED and X-ray obscuration is not strong.

In order to further examine the correspondence between optical properties, UV-optical-near-infrared color and morphology, and X-ray obscuration, the AGN were separated into 4 groups based on the restframe $u^*$-$H$ color and UV morphology. The restframe colors were calculated from the SED fitting results of LePhare. The morphology was inferred from the HSC $i$-band flux ratio between the HSC S20A $i$-band PSF and cmodel flux. The PSF (cmodel) flux is derived by fitting PSF (PSF or galaxy) model. If the AGN appears as a pointsource on the image then the ratio is expected to approach 1 while extended AGN will have a smaller flux ratio. We consider AGN with the flux-ratio larger than 0.95 as a point sources.

\begin{figure}[!htb]
    \centering
    \plotone{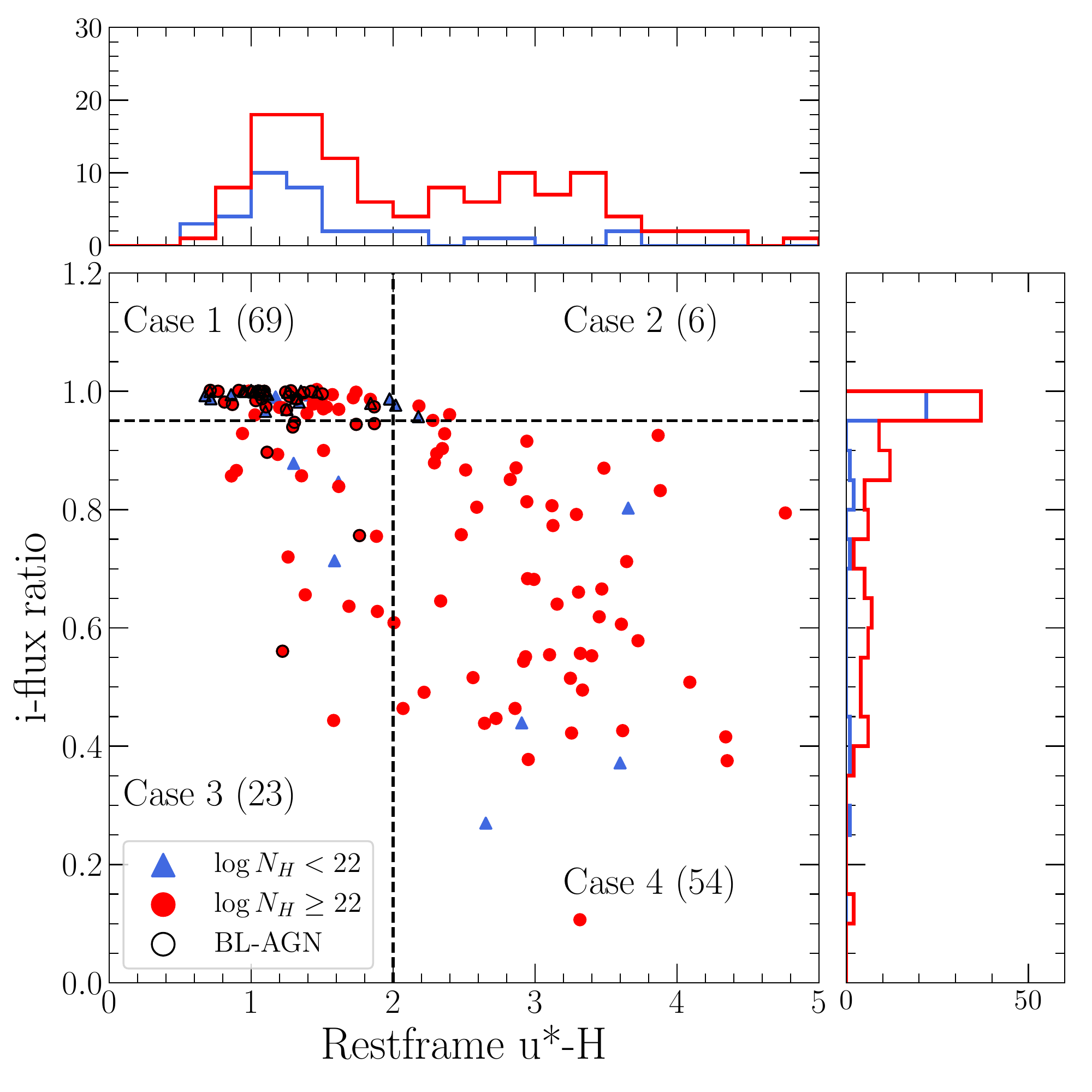}
    \caption{Rest-frame $u^*$-H colors vs. the HSC $i$-band flux-ratio. Red round and blue triangles symbols represent X-ray obscured and unobscured AGN respectively. Black-rimmed symbols are broad-line AGN. The distribution is separated into 4 quadrants based on the restframe color and morphology. The numbers in parentheses represent the number of objects in each quadrant. The side histograms show the projected morphology distribution (right) and color distribution (top).}
    \label{fig:uH_morph}
\end{figure}

The distribution of the high redshift AGN on the rest-frame $u^*$-$H$ color and the flux ratio plane is shown in Figure  \ref{fig:uH_morph}. Most of the X-ray unobscured AGN have blue rest-frame color and morphology similar to a point source. For the X-ray obscured AGN, approximately 49\% show extended morphology while the remaining sources possess morphology consistent with a point source. X-ray obscured AGN also show a broad color distribution where some X-ray obscured AGN have restframe optical near-infrared colors consistent with the X-ray unobscured AGN.

\begin{figure}[!htb]
    \centering
    \plotone{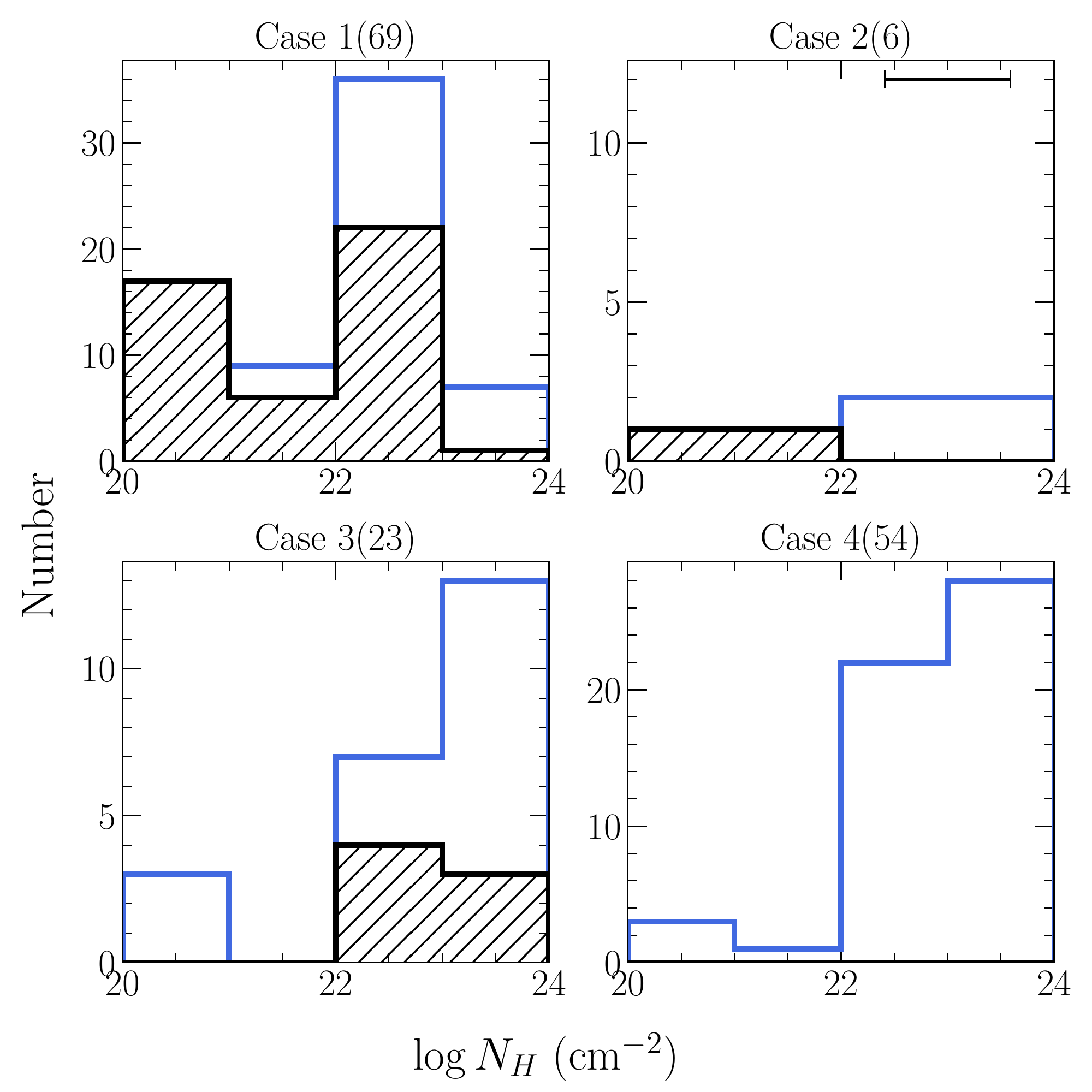}
    \caption{Observed $\log N_{\rm H}$ distribution of AGN in each color-morphology quadrant shown in blue. The $\log N_{\rm H}$ of known broad-line AGN are shown as black hatched histograms. The datapoint shows the typical $\log N_{\rm H}$ uncertainty. The number in parenthesis represent the number of objects in each quadrant.}
    \label{fig:nhdist_quad}
\end{figure}

The high-redshift AGN sample was divided into 4 samples based on the color against the morphology plane, and the column density distribution of each sample is shown in Figure \ref{fig:nhdist_quad}. Most of the AGN with extended morphology have $\log N_{\rm H}$ consistent with the X-ray obscured AGN with $\log N_{\rm H}\ (\rm  cm^{-2})>22$. On the other hand, blue point source AGN (case 1) has a mixture of column density $\log N_{\rm H}$ both consistent with X-ray obscured and unobscured AGN.

For each type of morphology, we performed a KS-test between the $\log N_{\rm H}$ distribution for those that are extended and those which are point source. The possibility that the blue and red point source (case 1 \& case 2) were drawn from the same parent distribution is rejected at a 0.01 level of significance ($D_{\rm max}=0.391$, $D_{crit}=0.277$). However, the number of samples compared between case 1 and case 2 is limited. On the other hand, the possibility that the blue and red extended sources (case 3 \& case 4) were drawn from the same parent distribution can not be rejected at a 0.01 level of significance ($D_{\rm max}=0.256$, $D_{crit}=0.405$). This suggests that both red and blue extend sources may contain X-ray obscured AGN.

The largest difference between the cumulative distribution of $\log N_{\rm H}$ of blue extended AGN and red extended AGN comes from the $\log N_{\rm H}\ (\rm  cm^{-2})=22-23$ bin. Obscuration of the largest column densities ($\log N_{\rm H}\ (\rm  cm^{-2})> 23$) generally occurs in the nuclear scale while moderate obscuration can occur on kpc scale \citep{2018ARA&A..56..625H}. It is possible that red extended X-ray obscured AGN have larger amounts of gas and dust in the host galaxy scale than blue extended X-ray obscured AGN. As a result, the red color can be explained by the larger dust extinction in the ISM. Another possibility is that the blue UV optical near-infrared colors found in blue extended X-ray obscured AGN is scattered light from the luminous quasars \citep{2013MNRAS.435.3306A,2016ApJ...819..111A,2018MNRAS.479.4936A,2020ApJ...897..112A} or from ongoing unobscured star-formation in the host galaxy and that the AGN is obscured as suggested by the extended morphology of the AGN host galaxy.

In order to examine the relation between the UV spectral properties and X-ray obscuration, the optical spectra of 26 AGN with SDSS spectroscopic data were examined. The morphology of 21 of these AGN are consistent with a point source object. Three sources have flux ratios $>0.90$ close to the stellarity threshold. Of the 26 AGN, two AGN has flux ratios consistent with an extended source with one AGN showing narrow CIV emission line ($\sigma<1000\ \rm km \ s^{-1}$). Among these 26 AGN, 2 AGN show absorption associated with the CIV emission line. The remaining AGN have spectra consistent with optically unobscured AGN with broad UV emission lines. Broad absorption line quasars (BAL-QSO) are known to be X-ray obscured \citep{2011MNRAS.416.2792P,2010A&A...517A..47M,2017MNRAS.464.4586P,2010A&A...515A...2S}, therefore some of the X-ray obscured AGN with blue UV continuum and pointsource morphology may be BAL-QSOs. X-ray obscuration may also be due to warm or ionized absorbers with no dust \citep{2005A&A...432...15P,2014MNRAS.437.3550M}.  This is overall consistent with the concept of the unified model \citep{1993ARA&A..31..473A,1995PASP..107..803U}. Detection of broad emission lines and blue continuum suggests that the line of sight towards the BLR and accretion disk is unobscured by dust but may contain ionized or dust-free X-ray absorbers along the line of sight due to the strong UV radiation.

We conclude that the trend in which X-ray unobscured AGN have a flat SED, blue UV-Optical color, and point source morphology while X-ray obscured AGN have a reddened SED and extended morphology is present but the correspondence is not tight. The large variety in the SED shapes may be due to different types of X-ray absorbers, scattered AGN emission, and the variety of host galaxy star formation and dust content.

\subsection{How Obscured Quasars are Missed by Optical Color-selection} \label{sec:des:cselect}
A large number of high redshift unobscured quasars are selected based on their optical color and morphology in wide and deep optical imaging surveys. This technique enables us to select faint unobscured AGN but can miss obscured AGN. We examine the relationship between the high redshift X-ray selected AGN and those with the optical color and morphology criteria as used in \citet{2018PASJ...70S..34A} and \citet{2022A&A...658A.175P}. 

\begin{figure}[!htb]
    \centering
    \plotone{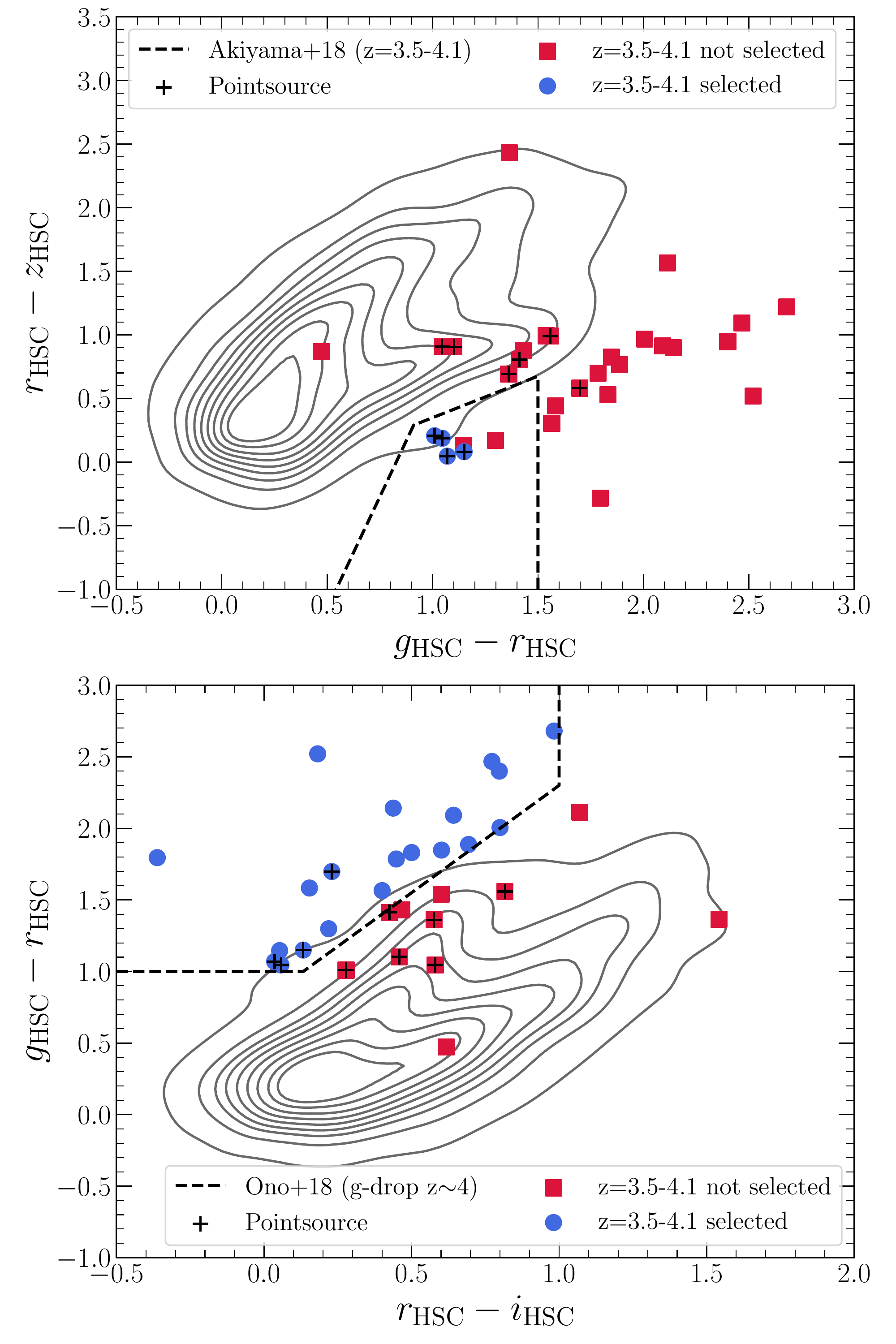}
    \caption{(Top) $g-r$ vs. $r-z$ color distribution. AGN at redshift 3.5-4.1 selected and not selected by the color morphology criterion of \citet{2018PASJ...70S..34A} is shown as blue round and red square symbols, respectively.  (Bottom) $r-i$ vs. $g-r$ color distribution. AGN at redshift  3.5-4.1 selected and not selected by the color criterion of  \citep{2018PASJ...70S..10O}  is shown as blue round and red square symbols, respectively. For both plots, the gray contours show the color distribution of AGN in the primary sample while objects with pointsource morphology are shown with crossed symbols.}
    \label{fig:cselect}
\end{figure}

Figure \ref{fig:cselect} shows the color distribution of AGN in the entire X-ray sample compared with those at redshift 3.5-4.1. The color and morphology selection in \citet{2018PASJ...70S..34A} was constructed to select AGN with point-source morphology at redshift 3.5 to 4.1. The color selection was designed to minimize contamination by AGN at other redshifts, galaxies at $z\sim1$, and low-mass galactic stars, which are the major contaminants. In this section, point source morphology is defined based on the adaptive moment measurement \citep{2003MNRAS.343..459H} used in \citet{2018PASJ...70S..34A} which is available in the HSC-SSP database. We adopt the condition $$\rm \frac{i\_hsmsourcemomentsround\_shape11}{i\_hsmpsfmoments\_shape11}<1.1 $$ $$\rm \frac{i\_hsmsourcemomentsround\_shape22}{i\_hsmpsfmoments\_shape22}<1.1$$ to define pointsource objects. For comparison with the pointsource morphology defined based on the flux ratio in Section \ref{sec:des:sed}, the adaptive moment criteria correspond to objects with flux ratios of $\sim0.97$. Approximately 16\% of AGN at redshift 3.5-4.1 were selected based on color selection criterion and half of them have pointsource morphology. It should be noted that the color criteria are determined for stellar objects brighter than $i<24$ mag, and most of the X-ray selected objects are fainter than $i>24$ mag.
 
In addition to a modified version of the color selection used in \citet{2018PASJ...70S..34A}, \citet{2022A&A...658A.175P} uses the Lyman-break criteria \citep{2018PASJ...70S..10O}. The Lyman-break criterion selects 65\% of the X-ray selected AGN at redshift 3.5 to 4.1 including both point source and extended objects.  The remaining objects which were not selected by the color selection criteria possess redder colors. We conclude that the AGN selection based on the optical color can miss obscured AGN which have reddened or host-dominated colors. Moreover, the application of morphological selection excludes obscured AGN which resides in extended host galaxies.

\section{Summary}

We construct a multiwavelength PSF-convolved photometric catalog from 12 deep and wide imaging datasets covering from $u^*$-band to 4.5$\rm \mu \mathrm{m}$ in the HSC-Deep XMM-LSS survey area using HSC $r$-band as the high-resolution prior. A sample of high redshift AGN was constructed by matching the XMM-SERVS X-ray point-source catalog \citep{2018MNRAS.478.2132C} with the multiwavelength catalog and selecting AGN above redshift 2 based on the best available redshift.  Thanks to the deep optical/NIR imaging, high photometric completeness was achieved and the AGN properties were examined with spectroscopic or photometric redshifts.  We perform a maximum-likelihood fitting assuming a modified absorption function of \citet{2014ApJ...786..104U} taking into account the survey bias against obscured AGN.  Based on the best-fit parameters, we obtain the following key results.

\begin{enumerate}
    \item We estimate that $76_{-3}^{+4}\%$ of high-redshift luminous quasars ($\log L_X\ (\rm erg\ s^{-1})>44.5$ \& $z>2$) are obscured. In the luminosity range of $\log L_X\ (\rm erg\ s^{-1})=44-45$ the obscured fraction is $85_{-3}^{+4}\%$ for $\log N_{\rm H}\ (\rm cm^{-2}) >22$ and $50_{-6}^{+7}\%$ for $\log N_{\rm H}\ (\rm cm^{-2}) >23$. The obscured fraction is consistent with those determined in the CDF-S in the same redshift range but larger than in previous studies (Section \ref{sec:res:intobs}).
    
    \item The obscured fraction above $z>2$ is larger than that in the local universe, consistent with the previous studies which suggest an increasing trend of the obscured fraction towards high redshift (Section \ref{sec:des:obsf}).
    
    \item The large obscured fraction can be explained with a model in which the obscured fraction continues to increase beyond redshift 2 and suggests that the obscured fraction of luminous AGN saturates at a higher redshift than that of less luminous AGN. Due to the saturation of the obscured fraction, the decreasing trend of the obscured fraction with luminosity could disappear at a high redshift ($z>3$, eg. \citealt{2014MNRAS.445.3557V,2018MNRAS.473.2378V}) (Section \ref{sec:des:obsf}).
    
    \item The large obscured fraction at high redshift may be a result of a large merger fraction at high redshift or due to large gas fractions in the circumnuclear and host galaxy. Due to the abundance of metals, gas, and dust, AGN feedback may be less efficient in clearing sight lines towards the nuclear region (Section \ref{sec:des:imp}). 
    
    \item  The trend in which X-ray unobscured AGN have pointsource morphology and blue-flat SEDs while X-ray obscured AGN have extended morphology and red SEDs is observed. However, the SED shows a large scatter in both cases. This suggests the correspondence is not strong. This may be a result of dust-free X-ray extinction, scattered AGN light, or unobscured star formation in obscured AGN host galaxies (Section \ref{sec:des:sed}). 

\end{enumerate}

 Based on the current evidence, the early phase of cosmological black hole growth is likely to occur in a highly obscured manner. Tracers of AGN activity that are strong against obscuration such as X-ray and mid-infrared emission will be key in tracing early SMBH growth as well as understanding the physical processes behind obscuration. Large legacy deep fields and future X-ray facilities such as. Athena, and Lynx, and infrared facilities such as the James Webb Space Telescope, will play an important role in enlarging obscured AGN samples and unveiling the physics of AGN obscuration at high redshift.

\vspace{+10pt}
\section*{acknowledgments}
The authors would like to thank the anonymous reviewers for the helpful comments which greatly improved the manuscript. In addition, the authors would like to thank Drs. Emiliano Merlin, Kohei Ichikawa, and Mitsuru Kokubo for the helpful discussions. 

The Hyper Suprime-Cam (HSC) collaboration includes the astronomical communities of Japan and Taiwan, and Princeton University.  The HSC instrumentation and software were developed by the National Astronomical Observatory of Japan (NAOJ), the Kavli Institute for the Physics and Mathematics of the Universe (Kavli IPMU), the University of Tokyo, the High Energy Accelerator Research Organization (KEK), the Academia Sinica Institute for Astronomy and Astrophysics in Taiwan (ASIAA), and Princeton University.  Funding was contributed by the FIRST program from the Japanese Cabinet Office, the Ministry of Education, Culture, Sports, Science and Technology (MEXT), the Japan Society for the Promotion of Science (JSPS), Japan Science and Technology Agency  (JST), the Toray Science  Foundation, NAOJ, Kavli IPMU, KEK, ASIAA, and Princeton University.

This paper is based [in part] on data collected at the Subaru Telescope and retrieved from the HSC data archive system, which is operated by Subaru Telescope and Astronomy Data Center (ADC) at NAOJ. Data analysis was in part carried out with the cooperation of Center for Computational Astrophysics (CfCA) at NAOJ.  We are honored and grateful for the opportunity of observing the Universe from Mauna kea, which has the cultural, historical and natural significance in Hawaii.

This paper makes use of software developed for Vera C. Rubin Observatory. We thank the Rubin Observatory for making their code available as free software at http://pipelines.lsst.io/. 

The Pan-STARRS1 Surveys (PS1) and the PS1 public science archive have been made possible through contributions by the Institute for Astronomy, the University of Hawaii, the Pan-STARRS Project Office, the Max Planck Society and its participating institutes, the Max Planck Institute for Astronomy, Heidelberg, and the Max Planck Institute for Extraterrestrial Physics, Garching, The Johns Hopkins University, Durham University, the University of Edinburgh, the Queen’s University Belfast, the Harvard-Smithsonian Center for Astrophysics, the Las Cumbres Observatory Global Telescope Network Incorporated, the National Central University of Taiwan, the Space Telescope Science Institute, the National Aeronautics and Space Administration under grant No. NNX08AR22G issued through the Planetary Science Division of the NASA Science Mission Directorate, the National Science Foundation grant No. AST-1238877, the University of Maryland, Eotvos Lorand University (ELTE), the Los Alamos National Laboratory, and the Gordon and Betty Moore Foundation.

These data were obtained and processed as part of the CFHT Large Area U-band Deep Survey (CLAUDS), which is a collaboration between astronomers from Canada, France, and China described in Sawicki et al. (2019, [MNRAS 489, 5202]).  CLAUDS is based on observations obtained with MegaPrime/ MegaCam, a joint project of CFHT and CEA/DAPNIA, at the CFHT which is operated by the National Research Council (NRC) of Canada, the Institut National des Science de l’Univers of the Centre National de la Recherche Scientifique (CNRS) of France, and the University of Hawaii. CLAUDS uses data obtained in part through the Telescope Access Program (TAP), which has been funded by the National Astronomical Observatories, Chinese Academy of Sciences, and the Special Fund for Astronomy from the Ministry of Finance of China. CLAUDS uses data products from TERAPIX and the Canadian Astronomy Data Centre (CADC) and was carried out using resources from Compute Canada and Canadian Advanced Network For Astrophysical Research (CANFAR).

Based on data obtained from the ESO Science Archive Facility with DOI(s): 	https://doi.org/10.18727/archive/58.

This research has made use of the NASA/IPAC Infrared Science Archive, which is funded by the National Aeronautics and Space Administration and operated by the California Institute of Technology.

This publication makes use of data products from the Two Micron All Sky Survey, which is a joint project of the University of Massachusetts and the Infrared Processing and Analysis Center/California Institute of Technology, funded by the National Aeronautics and Space Administration and the National Science Foundation.

This research made use of Photutils, an Astropy package for detection and photometry of astronomical sources (\citealt{larry_bradley_2020_4044744}).

\facilities{IRSA,CFHT, Subaru, VISTA, Spitzer, XMM-Newton}
\software{NumPy, SciPy, AstroPy, Matplotlib, Photutils, LePhare, SExtractor, SWarp, T-PHOT, PyPher}


\appendix
\section{Catalog Description} \label{appendix}

The description of the X-ray AGN catalog as well as an example is shown here in table \ref{tab:xagn_agntab_des} and \ref{tab:xagn_agntab_exp}, respectively. 

\startlongtable
\begin{deluxetable*}{llll}  \label{tab:xagn_agntab_des}
\tabletypesize{\footnotesize}
\tablecaption{X-ray AGN catalog Description}
\tablehead{\colhead{Number} & \colhead{Column Name} & \colhead{Unit} & \colhead{Description}}
\startdata
\multicolumn{4}{c}{X-ray Source Information (\citealt{2018MNRAS.478.2132C})}  \\
\hline
1 & Namexid & --- & X-ray Source ID \\
2 & RAxcen & deg & X-ray center right ascension \\
3 & DExcen & deg & X-ray center declination \\
4 & SBml & --- & 0.5-2 keV detection likelihood \\
5 & HBml & --- & 2-10 keV detection likelihood \\
6 & FBml & --- & 0.5-10 keV detection likelihood \\
7 & oircat & --- & Matched optical-IR Catalog \\
8 & RAoir & deg & OIR right ascension \\
9 & DEoir & deg & OIR declination \\
\hline 
\multicolumn{4}{c}{Additional Derived X-ray Information} \\
\hline
10 & SBcrt & ct/ks & 0.5-2 kev count-rate \tablenotemark{ab} \\
11 & e\_SBcrt & ct/ks & 0.5-2 kev count-rate uncertainty \tablenotemark{ab}  \\
12 & HBcrt & ct/ks & 2-10 kev count-rate \tablenotemark{ab}  \\
13 & e\_HBcrt & ct/ks & 2-10 kev count-rate uncertainty \tablenotemark{ab}  \\
14 & HR & --- & Hardness ratio \tablenotemark{b}  \\
15 & e\_HR & --- & Hardness ratio uncertainty \tablenotemark{b}  \\
16 & blagn & --- & Optical Broadline AGN \tablenotemark{c}  \\
\hline 
\multicolumn{4}{c}{Optical and IR PSF-convolved Photometry} \\
\hline
17 & Namexhsc & --- & XHSC source identifier \\
18 & RAxhsc & deg & XHSC source right ascension \\
19 & DExhsc & deg & XHSC source declination \\
20 & tract & --- & HSC tract identifier \\
21 & patch & --- & HSC patch identifier \\
22 & nircat & --- & Detection flag \tablenotemark{d} \\
23 & usmag & mag & CLAUDS us-band magnitudes \\
24 & e\_usmag & mag & CLAUDS us-band magnitudes uncertainty \\
25 & gmag & mag & HSC g-band magnitudes \\
26 & e\_gmag & mag & HSC g-band magnitudes uncertainty \\
27 & rmag & mag & HSC r-band magnitudes \\
28 & e\_rmag & mag & HSC r-band magnitudes uncertainty \\
29 & imag & mag & HSC i-band magnitudes \\
30 & e\_imag & mag & HSC i-band magnitudes uncertainty \\
31 & zmag & mag & HSC z-band magnitudes \\
32 & e\_zmag & mag & HSC z-band magnitudes uncertainty \\
33 & ymag & mag & HSC y-band magnitudes \\
34 & e\_ymag & mag & HSC y-band magnitudes uncertainty \\
35 & Ymag & mag & VIDEO Y-band magnitudes \\
36 & e\_Ymag & mag & VIDEO Y-band magnitudes uncertainty \\
37 & Jmag & mag & VIDEO J-band magnitudes \\
38 & e\_Jmag & mag & VIDEO J-band magnitudes uncertainty \\
39 & Hmag & mag & VIDEO H-band magnitudes \\
40 & e\_Hmag & mag & VIDEO H-band magnitudes uncertainty \\
41 & Ksmag & mag & VIDEO Ks-band magnitudes \\
42 & e\_Ksmag & mag & VIDEO Ks-band magnitudes uncertainty \\
43 & ch1mag & mag & SERVS 3.6-um magnitudes \\
44 & e\_ch1mag & mag & SERVS 3.6-um magnitudes uncertainty \\
45 & ch2mag & mag & SERVS 4.5-um magnitudes \\
46 & e\_ch2mag & mag & SERVS 4.5-um magnitudes uncertainty \\
47 & EBV & mag & E(B-V) \\
48 & AV & mag & V-band Attenuation \\
49 & f\_usmag & --- & CLAUDS us-band bad photometry flag \tablenotemark{e}  \\
50 & f\_gmag & --- & HSC g-band bad photometry flag \tablenotemark{e} \\
51 & f\_rmag & --- & HSC r-band bad photometry flag \tablenotemark{e} \\
52 & f\_imag & --- & HSC i-band bad photometry flag \tablenotemark{e} \\
53 & f\_zmag & --- & HSC z-band bad photometry flag \tablenotemark{e} \\
54 & f\_ymag & --- & HSC y-band bad photometry flag \tablenotemark{e} \\
55 & f\_Ymag & --- & VIDEO Y-band bad photometry flag \tablenotemark{e} \\
56 & f\_Jmag & --- & VIDEO J-band bad photometry flag \tablenotemark{e} \\
57 & f\_Hmag & --- & VIDEO H-band bad photometry flag \tablenotemark{e} \\
58 & f\_Ksmag & --- & VIDEO Ks-band bad photometry flag \tablenotemark{e} \\
59 & f\_ch1mag & --- & SERVS 3.6-um bad photometry flag \tablenotemark{e} \\
60 & f\_ch2mag & --- & SERVS 4.5-um bad photometry flag \tablenotemark{e} \\
61 & f\_leda & --- & LEDA Association Flag \tablenotemark{f} \\
\hline 
\multicolumn{4}{c}{Spectroscopic Redshift \& LePhare Photo-z} \\
\hline
62 & zspecid & --- & Spectroscopic redshift ID \\
63 & zspec & --- & Spectroscopic redshift \\
64 & zphot & --- & Best photometric redshift \\
65 & lzphot & --- & Upper 68\% Confidence Photo-z \\
66 & uzphot & --- & Lower 68\% Confidence Photo-z \\
67 & chibest & --- & Best-fit Chi-square \\
68 & nband & --- & Number of bands used \\
69 & zsec & --- & Secondary photo-z \\
70 & chisec & --- & Secondary photo-z Chi-square \\
71 & chistar & --- & Galactic star photo-z Chi-square \\
72 & usMag & mag & CLAUDS us-band absolute magnitudes \tablenotemark{g} \\
73 & gMag & mag & HSC g-band absolute magnitudes \tablenotemark{g} \\
74 & rMag & mag & HSC r-band absolute magnitudes \tablenotemark{g} \\
75 & iMag & mag & HSC i-band absolute magnitudes \tablenotemark{g} \\
76 & zMag & mag & HSC z-band absolute magnitudes \tablenotemark{g} \\
77 & yMag & mag & HSC y-band absolute magnitudes \tablenotemark{g} \\
78 & YMag & mag & VIDEO Y-band absolute magnitudes \tablenotemark{g} \\
79 & JMag & mag & VIDEO J-band absolute magnitudes \tablenotemark{g} \\
80 & HMag & mag & VIDEO H-band absolute magnitudes \tablenotemark{g} \\
81 & KsMag & mag & VIDEO Ks-band absolute magnitudes \tablenotemark{g} \\
82 & ch1Mag & mag & SERVS 3.6-um absolute magnitudes \tablenotemark{g} \\
83 & ch2Mag & mag & SERVS 4.5-um absolute magnitudes \tablenotemark{g} \\
\hline 
\multicolumn{4}{c}{Derived AGN Properties Used in The Analysis} \\
\hline
84 & zprime & --- & Best-redshift used in analysis \\
85 & SBflux & $\rm 10^{-3} Wm^{-2}$ & 0.5-2 keV Band flux \\
86 & e\_SBflux & $\rm 10^{-3} Wm^{-2}$ & 0.5-2 keV Band flux uncertainty \\
87 & HBflux & $\rm 10^{-3} Wm^{-2}$ & 2-10 keV Band flux \\
88 & e\_HBflux & $\rm 10^{-3} Wm^{-2}$ & 2-10 keV Band flux uncertainty \\
89 & FBflux & $\rm 10^{-3} Wm^{-2}$ & 0.5-10 keV Band flux \\
90 & eFBflux & $\rm 10^{-3} Wm^{-2}$ & 0.5-10 keV Band flux uncertainty \\
91 & logNH & [$\rm cm^{-2}$] & Hydrogen column density \\
92 & llogNH & [$\rm cm^{-2}$] & Column density lower limit \\
93 & ulogNH & [$\rm cm^{-2}$] & Column density lower limit \\
94 & SBloglx & [$\rm 10^{-7}W$] & 2-10 keV luminosity from SB \tablenotemark{a} \\
95 & e\_SBloglx & [$\rm 10^{-7}W$] & 2-10 keV luminosity uncertainty from SB \tablenotemark{a}  \\
96 & HBloglx & [$\rm 10^{-7}W$] & 2-10 keV absorption corrected luminosity from HB \tablenotemark{a}  \\
97 & e\_HBloglx & [$\rm 10^{-7}W$] & 2-10 keV absorption corrected luminosity uncertainty HB \tablenotemark{a}  \\
98 & FBloglx & [$\rm 10^{-7}W$] & 2-10 keV absorption corrected luminosity from FB \tablenotemark{a}  \\
99 & e\_FBloglx & [$\rm 10^{-7}W$] & 2-10 keV absorption corrected luminosity uncertainty from FB \tablenotemark{a} 
\enddata
\tablenotetext{a}{SB=0.5-2 keV, HB=2-10 keV, FB=0.5-10 keV}
\tablenotetext{b}{PN-equivalent count-rates}
\tablenotetext{c}{BLAGN flag from \citet{2018MNRAS.478.2132C}}
\tablenotetext{d}{0=r-band detected 1=H-band detected}
\tablenotetext{e}{0=Clean photometry 1=Bad photometry}
\tablenotetext{f}{0/1=Not associated/Associated with sources in HyperLeda catalog.}
\tablenotetext{g}{Absolute magnitudes calculated from LePhare}
\end{deluxetable*}

\begin{deluxetable}{ccccccccccc} \label{tab:xagn_agntab_exp}
\tablecaption{X-ray AGN Catalog Table Example} 
\tablehead{\colhead{Namexid} & \colhead{RAxhsc} & \colhead{DExhsc} & \colhead{imag} &  \colhead{e\_imag} & \colhead{HBml} & \colhead{zspec} & \colhead{zphot} & \colhead{logNH} & \colhead{HBloglx} & \colhead{e\_HBloglx }\\ \colhead{ } & \colhead{$\mathrm{deg}$} & \colhead{$\mathrm{deg}$} & \colhead{$\mathrm{mag}$} & \colhead{$\mathrm{mag}$} & \colhead{ } & \colhead{ } & \colhead{ } & \colhead{dex(1 / cm2)} & \colhead{[10-7W]} & \colhead{[10-7W]}}
\startdata
XMM00045 & 34.2305158 & -5.393255 & 21.742 & 0.01 & 26.4 & 2.2184 & 2.2464 & 21.79 & 44.264 & 0.085 \\
XMM00113 & 34.2661121 & -5.5280322 & 21.05 & 0.01 & 9.1 & 2.2815 & 0.0 & 22.51 & 44.562 & 0.086 \\
XMM00134 & 34.2740016 & -4.9119973 & 25.285 & 0.022 & 13.5 & 2.797 & 2.2013 & 21.89 & 44.563 & 0.102 \\
XMM00145 & 34.2813749 & -4.5672507 & 25.309 & 0.022 & 54.6 &  & 3.1306 & 23.33 & 44.733 & 0.084 \\
XMM00201 & 34.3171034 & -5.1264044 & 22.789 & 0.01 & 31.2 & 2.3313 & 2.149 & 20.0 & 44.549 & 0.081
\enddata
\tablecomments{Table \ref{tab:xagn_agntab_exp} is published in its entirety in the machine-readable format. A portion is shown here for guidance regarding its form and content.}
\end{deluxetable}

\bibliography{ref}
\bibliographystyle{aasjournal}




\end{document}